\documentclass[11pt,oneside]{article}

\usepackage[margin=1in]{geometry}
\usepackage{palatino}
\usepackage{url}
\usepackage{textcomp}

\usepackage{color,xcolor,soul} 
\usepackage{setspace}          
\usepackage{framed}            
\usepackage{subcaption}
\usepackage{graphicx,tikz}
\usepackage{amsmath,amssymb,amsfonts}
\usepackage{mathtools}         
\usepackage{natbib}
\usepackage[colorlinks,citecolor=blue]{hyperref}
\usepackage[capitalize]{cleveref} 
\usepackage{algpseudocode,algorithmicx,algorithm}

\usepackage[colorlinks, citecolor=blue]{hyperref}  

\newcounter{lemma}
\newenvironment{lemma}{\refstepcounter{lemma}%
\noindent\textit{Lemma \thelemma.}\ \em\rmfamily}{}

\newcounter{claim}
\newenvironment{claim}{\refstepcounter{claim}%
\noindent\textit{Claim \theclaim.}\ \em\rmfamily}{}

\newcounter{proposition}
\newenvironment{proposition}{\refstepcounter{proposition}%
\noindent\textit{Proposition \theproposition.}\ \em\rmfamily}{}

\newenvironment{proposition*}[1]{\refstepcounter{proposition}%
\noindent\textit{Proposition \theproposition~(#1)}\ \em\rmfamily}{}

\newcounter{theorem}
\newenvironment{theorem}{\refstepcounter{theorem}%
\noindent\textit{Theorem \thetheorem.}\ \em\rmfamily}{}

\newcounter{corollary}

\newcounter{definition}
\newenvironment{definition*}[1]{\refstepcounter{definition}%
\noindent\textit{Definition \thedefinition~(#1)}\ \rmfamily}{}

\newcounter{problem}

\newcounter{assumption}
\newenvironment{assumption*}[1]{\refstepcounter{assumption}%
\noindent\textit{Assumption \theassumption~(#1)}\ \rmfamily}{}
\newenvironment{assumption}{\refstepcounter{assumption}%
\noindent\textit{Assumption \theassumption.}\ \em\rmfamily}{}

\newcounter{remark}
\newenvironment{remark}{\refstepcounter{remark}%
\noindent\textit{Remark \theremark.}\ \em\rmfamily}{}

\newcounter{example}
\newenvironment{example}{\refstepcounter{example}%
\noindent\textit{Example \theexample.}\ \em\rmfamily}{}

\newenvironment{myproof}{\indent\textit{Proof.}\ \rmfamily}{\hfill$\square$}

\DeclareMathOperator*{\argmax}{arg\,max}
\newcommand{\be}{\begin{equation}}
\newcommand{\ee}{\end{equation}}
\newcommand{\E}{\mathrm{E}}
\newcommand{\epi}{\overline{\pi}}
\newcommand{\cpi}{\widehat{\pi}}
\newcommand{\Pieps}{\Pi_{\epsilon}^{\mathrm{eq}}}
\newcommand{\ePieps}{\overline{\Pi}_{\epsilon}^{\mathrm{eq}}}
\newcommand{\ePiepsv}{\overline{\Pi}_{\epsilon,\nu}^{\mathrm{eq}}}
\newcommand{\eU}{\overline{U}}
\newcommand{\SW}{\mathrm{SW}}
\newcommand{\eSW}{\overline{\mathrm{SW}}}
\newcommand{\eSWv}{\overline{\mathrm{SW}}_\nu}
\newcommand{\explore}{\mathcal{E}}
\newcommand{\taue}{\tau_{\mathrm{explore}}}
\newcommand{\ideal}{\mathrm{ideal}}
\newcommand{\wtheta}{\widehat{\theta}}
\newcommand{\TV}{d_\mathrm{TV}}
\newcommand{\lambdabar}{\overline{\lambda}}
\newcommand{\wlambda}{\widehat{\lambda}}
\newcommand{\floor}[1]{\left\lfloor #1 \right\rfloor}

\begin{document}

\title{Decentralized Optimal Equilibrium Learning in Stochastic Games via Single-bit Feedback}

\author{
Seref Taha Kiremitci\thanks{S. T. Kiremitci and M. O. Sayin are with the Department of Electrical and Electronics Engineering, Bilkent University, Ankara, T\"{u}rkiye. Emails: taha.kiremitci@bilkent.edu.tr, sayin@ee.bilkent.edu.tr.}%
\and
Ahmed Said Donmez\thanks{A. S. Donmez was with the Department of Electrical and Electronics Engineering, Bilkent University, Ankara, T\"{u}rkiye. Email: said.donmez@bilkent.edu.tr.}%
\and
Muhammed O. Sayin\footnotemark[1]%
}

\date{}
\maketitle

\begin{abstract}
We study decentralized equilibrium selection in stochastic games under severe information and communication constraints. In such settings, convergence to equilibrium alone is insufficient, as stochastic games typically admit many equilibria with markedly different welfare properties. We address decentralized optimal equilibrium selection, where agents coordinate on equilibria that optimize a designer-specified social welfare objective while allowing heterogeneous tolerance to deviations from strict best responses. Agents observe only the global state trajectory and their realized rewards, and exchange a single randomized bit of feedback per agent per round. This semantic content/discontent signaling mechanism implicitly aligns decentralized learning dynamics with the global welfare objective. We develop explore-and-commit and online variants applicable to general stochastic games, accommodating heterogeneous model-based or model-free methods for solving the induced Markov decision processes, and establish explicit finite-time regret guarantees, showing logarithmic expected regret under mild conditions.
\end{abstract}


\section{Introduction}
\label{sec:introduction}
Decentralized multi-agent systems increasingly operate in environments where agents interact strategically over time without centralized coordination, such as networked control systems \citep{ref:Yuksel13, ref:Zhang21, ref:Ozdaglar21}. In these settings, agents repeatedly make decisions that affect both immediate rewards and future system evolution, and their interactions are naturally modeled as stochastic (Markov) games. The combination of strategic interaction, intertemporal decision-making, partial observability, and limited communication makes principled coordination a fundamental challenge.

A central difficulty in such systems is equilibrium selection. While much of the decentralized learning literature (reviewed in Subsection \ref{sec:lit} later) focuses on convergence to equilibrium, equilibrium convergence alone does not address which equilibrium is reached, nor whether the resulting outcome is socially desirable. This issue is particularly acute in stochastic games, where equilibrium sets are typically large and unstructured, and where different equilibria can lead to markedly different outcomes in terms of efficiency, fairness, and long-run performance. Moreover, many system-level objectives of practical interest go beyond additive welfare, encompassing nonlinear and fairness-sensitive criteria, such as proportional-fair or product-based objectives, that cannot be reduced to local payoff maximization or standard potential-game formulations, especially in stochastic games.

At the same time, requiring agents to follow exact best responses may be overly restrictive. In many decentralized systems, agents can tolerate deviations from strict equilibrium behavior if such deviations improve collective performance, and this tolerance may vary across agents due to incentives, bounded rationality, or design considerations. This introduces a fundamental tradeoff between equilibrium stability and social welfare maximization, which is largely unexplored in decentralized learning for stochastic games and is especially challenging when agents observe only local outcomes and must reason over state-dependent, intertemporal effects.

This paper addresses decentralized optimal equilibrium selection in general finite discounted stochastic games. Our objective is not merely to reach an equilibrium, but to enable decentralized agents to optimize over the equilibrium set with respect to a designer-specified social welfare objective, while explicitly allowing heterogeneous tolerance to deviation from best responses. This perspective yields a unified framework that spans a continuum from strict non-cooperative equilibrium enforcement to more cooperative, welfare-oriented coordination, under severe information and communication constraints.

To this end, we introduce \textit{Decentralized Optimal Equilibrium Learning (DOEL)}, a learning framework in which agents rely only on local observations and exchange a single randomized bit of feedback per agent per round. The core mechanism is a semantic–randomized signaling scheme in which agents probabilistically signal “content” or “discontent” based on local value comparisons relative to agent-specific tolerance levels. Although each signal carries only one bit, the joint distribution of signals implicitly encodes the global social welfare objective. Under uniform exploration, jointly content-endorsed strategies are sampled according to an entropy-regularized distribution over equilibria, assigning the highest probability mass to equilibria that optimally balance welfare maximization and equilibrium stability.

Building on this mechanism, we develop decentralized learning dynamics that operate in general stochastic games and explicitly accommodate agent heterogeneity. Agents may differ both in their tolerance to suboptimal responses and in their computational preferences and capabilities, including the use of model-based or model-free methods to solve the induced Markov decision problems. We present an explore-and-commit scheme, as well as an online variant that interleaves exploration and exploitation, while preserving the same underlying coordination mechanism.

The proposed framework applies to general finite stochastic games and general social welfare objectives, without relying on special game structures, centralized training, or shared model information. For both the explore-and-commit scheme and its online counterpart, we establish explicit finite-time regret guarantees with respect to the welfare of the optimal equilibrium under the specified tolerance profile. In particular, under mild regularity conditions, the expected regret grows at most logarithmically with time, up to approximation terms induced by exploration parameters. These results demonstrate that extreme communication efficiency—one bit per agent per round—is compatible with principled control over the equilibrium–welfare tradeoff in dynamic strategic environments.

\subsection{Related Works}\label{sec:lit}

Nash equilibrium provides a stable solution concept under which no agent has an incentive to deviate unilaterally \citep{ref:Basar98}. Whether such equilibria can emerge through decentralized, non-equilibrium adaptation of learning agents has been a central question in the (algorithmic) learning-in-games literature, providing a behavioral and dynamical justification for the extensive study of equilibrium analysis and computation \citep{ref:Fudenberg09}.

Early work on learning dynamics focused on the repeated play of normal-form games. Notably, even rough best-response-type learning rules, such as fictitious play, have been shown to converge to equilibrium in important classes of games, including zero-sum games \citep{ref:Robinson51, ref:Harris98} and potential or identical-interest games \citep{ref:Monderer96}. At the same time, it is well understood that uncoupled learning dynamics cannot guarantee convergence to Nash equilibrium in general classes of games \citep{ref:Hart03}, despite the existence of mixed-strategy equilibria in all finite games \citep{ref:Basar98}. These results highlight the power and the limitations of decentralized learning in static environments and motivate extensions to richer dynamic interaction models.

Stochastic games, introduced by Shapley, generalize discounted Markov decision processes to non-cooperative multi-agent environments and capture inter-temporal strategic interactions \citep{ref:Shapley53}. Shapley established the existence of Markov stationary equilibrium in two-agent zero-sum stochastic games via minimax value iteration, and existence results were later extended to general-sum stochastic games \citep{ref:Fink64}. Building on these foundations, \citet{ref:Littman94} introduced minimax-Q learning as a model-free extension of Shapley’s minimax iteration for zero-sum stochastic games , while \citet{ref:Hu03} proposed Nash-Q learning for general-sum stochastic games, establishing convergence under restrictive assumptions.

More recently, stochastic games have become a central modeling framework for multi-agent reinforcement learning, driven in part by empirical successes in large-scale artificial intelligence applications \citep{ref:Silver17, ref:Mirowski18}. In parallel, significant theoretical effort has been devoted to the design and analysis of learning dynamics in stochastic games with provable convergence and performance guarantees under decentralized information structures \citep{ref:Zhang21, ref:Ozdaglar21}. These works underscore the intrinsic difficulty of learning in stochastic games even before considerations of efficiency or equilibrium selection.

Within this line of research, several works have established convergence guarantees for decentralized learning dynamics in structured classes of stochastic games. \citet{ref:Arslan17} proposed payoff-based learning dynamics that converge to Markov stationary pure-strategy equilibria in broad classes of stochastic teams and games. \citet{ref:Leslie20} extended continuous-time best-response dynamics to stochastic games, proving convergence in two-agent zero-sum settings via a two-timescale separation between value estimation and strategy adaptation. These works initiated a broader line of research on convergent best-response-type learning dynamics in stochastic games, including variants of fictitious play and Q-learning, for important subclasses such as zero-sum and identical-interest stochastic games \citep{ref:Sayin22a, ref:Sayin21, ref:Sayin22b, ref:Donmez26, ref:Baudin22,  ref:Chen23, ref:Maheshwari25}.

Gradient-based learning dynamics have also been studied extensively in stochastic games, providing an alternative algorithmic paradigm \citep{ref:Daskalakis20,  ref:Leonardos25,  ref:Zhang24}. Complementary work has focused on learning (coarse) correlated equilibrium notions in general-sum stochastic games under decentralized information \citep{ref:Jin21, ref:Mao23}. Across all these, the primary objective is convergence to a given equilibrium notion under increasingly general stochastic games. Questions of equilibrium selection and welfare optimization, however, are largely orthogonal to these approaches.

Indeed, while equilibrium provides stability against unilateral deviations, equilibria can be arbitrarily inefficient from a system-level perspective. Balancing equilibrium stability with social welfare is therefore a fundamental challenge in multi-agent systems. For decentralized payoff-sum maximization in static games, \citet{ref:Marden14} proposed payoff-based learning dynamics that are stochastically stable at Pareto-optimal solutions in general classes of interdependent games, and \citet{ref:Pradelski12} extended these ideas to Pareto-optimal equilibrium selection. While these establish asymptotic convergence guarantees in the sense of stochastic stability, they are limited to the repeated play of normal-form games and primarily address additive welfare objectives.

Related efficiency guarantees have been obtained in potential games, where logit-based learning dynamics converge to potential maximizers \citep{ref:Marden12, ref:Arslan04, ref:Tatarenko17} and have been extended to stochastic teams \citep{ref:Donmez25, ref:Yongacoglu22}. Moreover, certain payoff-sum maximization problems can be recast as potential games \citep{ref:Arslan07}. However, potential games constitute a restrictive class, their extension to stochastic games requires strong structural assumptions \citep{ref:Leonardos25}, and potential maximization does not generally coincide with maximizing a desired social welfare objective.

A separate line of work studies communication-based distributed learning algorithms for payoff-sum maximization in cooperative multi-agent systems, where networked information exchange mitigates the limitations of local observations \citep{ref:Nedic09, ref:Zhang21}. Considerable effort has also been devoted to reducing communication overhead through quantized or compressed message passing, including single-bit and sign-based schemes, while retaining convergence guarantees \citep{ref:Sayin13, ref:Sayin14, ref:Zhang19, ref:Chen11, ref:Cao23a, ref:Cao23b}. These approaches, however, focus on cooperative optimization of additive welfare objectives and do not account for strategic incentives. In contrast, our work employs communication for strategy coordination rather than forming consensus.

Finally, \citet{ref:Yang25} proposed a learning dynamic reaching equilibria maximizing the minimum transformed utility across agents in general finite games. While this addresses equilibrium selection under fairness considerations, it is restricted to static games and stochastic stability guarantees. A precursor to the present work introduced single-bit coordination dynamics for Pareto-efficient outcomes in finite games \citep{ref:Kiremitci26}. The present paper extends this line of research to stochastic games and equilibrium selection under general social welfare objectives, through both explore-and-commit and online learning variants with explicit performance guarantees.

The paper is organized as follows. In Sections \ref{sec:problem} and \ref{sec:DOEL}, we describe stochastic games and our decentralized optimal equilibrium learning dynamics, respectively. In Sections \ref{sec:results} and \ref{sec:proofs}, we present the main results and the proofs, respectively. We provide illustrative examples in Section \ref{sec:simulations} and concluding remarks in Section \ref{sec:conclusion}. Appendices \ref{app:SW}-\ref{app:sum} include the proofs of four technical lemmas.

\section{Problem Formulation}\label{sec:problem}

Consider an $n$-agent stochastic game characterized by the tuple 
$\langle S, \{A^i,r^i\}_{i\in N}, p, \gamma\rangle$, 
where $N:=\{1,\ldots,n\}$ is the index set of agents, $S$ is the \textit{finite} set of states, $A^i$ is the \textit{finite} set of actions for agent $i$, 
$r^i:S\times A \to \mathbb{R}$ with $A:=\prod_j A^j$ is $i$'s reward function, 
$p(s_+\mid s,a)$ denotes the transition kernel giving the probability that state $s$ 
transits to $s_+$ when the joint action $a\in A$ is taken, 
and $\gamma\in[0,1)$ is the discount factor. 

Each agent $i$ adopts a \textit{pure} Markov strategy $\pi^i:S\to A^i$, 
and $\Pi^i$ denotes the \textit{finite} set of such strategies. 
For a joint strategy $\pi:=(\pi^j)_{j=1}^n\in\Pi:=\prod_j\Pi^j$, 
the expected discounted return of agent $i$ is defined by
\be\label{eq:utility}
U^i(\pi^i,\pi^{-i})
:=\E\!\left[\sum_{t=0}^{\infty}\gamma^t r^i(s_t,a_t)\right],
\ee
where $\pi^{-i}:=(\pi^j)_{j\neq i}$ is the others' joint strategy, 
$(s_t,a_t)$ denotes the state–action pair at stage $t$, 
and the expectation is taken over the trajectory distribution induced by $p$.

As a solution concept, we focus on Nash equilibria:

\begin{definition*}{Equilibrium}
A strategy profile $\pi$ is a \emph{(Markov stationary pure-strategy) $\epsilon$-equilibrium} 
if, for each agent $i$,
\be\label{eq:best}
U^i(\pi^i,\pi^{-i})
\ge U^i(\tilde{\pi}^i,\pi^{-i})-\epsilon^i,
\quad \forall \tilde{\pi}^i\in\Pi^i,
\ee
for some $\epsilon^i\ge0$. 
It is an exact equilibrium provided that $\epsilon^i=0$ for all $i$. Given $\epsilon=(\epsilon^i)_{i=1}^n$, we define
\be\label{eq:bestset}
\Pieps := \left\{\pi\in \Pi: U^i(\pi) \geq \max_{\tilde{\pi}^i\in \Pi^i}U^i(\tilde{\pi}^i,\pi^{-i}) - \epsilon^i, \forall i\right\}
\ee
as the set of all such equilibria.\footnote{There exists a maximizer $\tilde{\pi}^i$ in \eqref{eq:bestset} since the problem reduces to a finite discounted Markov decision process given $\pi^{-i}$.}
\end{definition*}

We further focus on \textit{equilibrium selection}.

\begin{definition*}{Social Welfare}
Given the utility functions $(U^i(\cdot))_{i=1}^n$, the \textit{social welfare} is defined by
\be\label{eq:SW}
\SW(\pi):=\sum_{i=1}^n w^i\cdot g^i(U^i(\pi))
\ee
for some positive weights $w^i>0$ and a monotonically increasing transformation $g^i(\cdot)$.
\end{definition*}

\begin{example} \label{Ex:sum_product}
For weights $w^i=1$, the identity transformation $g^i(x)=x$ yields the maximization of the sum of utilities:
\be\label{eq:SWsum}
\SW_\mathrm{Sum}(\pi) = \sum_{i=1}^n U^i(\pi).
\ee
For positive utilities, the logarithm function $g^i(x)=\log(x)$ yields the maximization of the product of utilities:
\be\label{eq:SWproduct}
\SW_\mathrm{Product}(\pi) = \sum_{i=1}^n \log(U^i(\pi)) = \log\prod_{i=1}^n U^i(\pi).
\ee
The heterogenous weights $(w^i)_{i=1}^n$ determine the share of the local utilities in the social welfare.
\end{example}

\begin{definition*}{Optimal Equilibrium}
An equilibrium $\pi_\ast\in\Pieps$ is an 
\emph{optimal $\epsilon$-equilibrium} if it maximizes the 
social welfare:
\be\label{eq:optimal}
\pi_\ast\in
\argmax_{\pi\in\Pieps}
\left\{\SW(\pi)\right\}.
\ee
\end{definition*}

Heterogeneous tolerance levels $(\epsilon^i)_{i=1}^n$ 
capture different sensitivities to suboptimal responses, 
ranging from full non-cooperativeness ($\epsilon^i\rightarrow0$) 
to full cooperativeness ($\epsilon^i\!\to\!\infty$).  

\begin{remark} \label{remark1}
The $\epsilon$-equilibrium set is finite as $\Pieps\subseteq \Pi$, which ensures the existence of a maximizer in \eqref{eq:optimal} if it is non-empty. Pure-strategy $\epsilon$-equilibrium does not necessarily exist in every game for arbitrarily small tolerance levels $\epsilon$ and $\Pieps$ might be an empty set for certain $\epsilon$. However, there always exist sufficiently large tolerance levels yielding $\Pieps \neq \varnothing$, e.g., $\Pieps= \Pi$. 
\end{remark}

We consider scenarios in which agents do \emph{not} know the model 
$\langle S,\{A^i,r^i\}_{i=1}^n,p,\gamma\rangle$ yet they know their $w^i$ and $g^i(\cdot)$, determining their share in the social welfare. Each agent observes only its local actions and realized rewards. 
Our goal is to develop decentralized learning dynamics through which 
heterogeneous agents can reach an optimal (approximate) equilibrium 
using only local information and \emph{single-bit} feedback. We focus on \emph{regret-based} analysis to evaluate effectiveness. Let $\pi_t$ denote the strategy profile at time $t$. Given a time budget $T$, we define the cumulative regret of the agents’ joint play relative to the optimal welfare in~\eqref{eq:optimal} as
\begin{equation}\label{eq:regret}
R_T :=
T\!\max_{\pi\in\Pieps}
\left\{\SW(\pi)\right\}
 - \sum_{t=0}^{T-1}\SW(\pi_t).
\end{equation}

\section{Decentralized Optimal Equilibrium Learning}\label{sec:DOEL}

Decentralized learning of optimal equilibrium poses several challenges:
\begin{itemize} 
\item The feasible set $\Pieps$ generally lacks a tractable structure, making the optimization problem~\eqref{eq:optimal} computationally prohibitive beyond exhaustive search even under centralized computation with full model knowledge. 
\item The absence of model information and the decentralized nature of the agents further exacerbate this challenge.
\end{itemize}
In the following, we present decentralized optimal equilibrium learning (DOEL) dynamics addressing these challenges in explore-and-commit and online schemes through a single-bit feedback mechanism.

\subsection{Single-bit Feedback Mechanism}\label{sec:signaling}

We propose a decentralized learning scheme that relies on minimal communication: each agent broadcasts and receives only 
\emph{single-bit feedback} signals when they are exploring their strategies.  Distinctly, our single-bit signaling scheme is not about diffusing local information among agents so that they reach a consensus on the global objective while compressing the information shared, e.g., via quantization, for minimal communication load. 
Instead, it leverages a \emph{probabilistic signaling mechanism} that implicitly 
aligns local strategies with the global objective  by leveraging independent randomization. 

We let each agent $i$ broadcast a one-bit message indicating \textit{content} `$1$' or \textit{discontent} `$0$' to all others. 
Since the feasibility constraint \eqref{eq:best} defining $\Pieps$ can be decomposed across agents, given a joint strategy $\pi$ and the corresponding utility $U^i(\pi)$, agent $i$ can \textit{independently} signal 
\be\label{eq:signal}
m^i = \left\{\begin{array}{ll}
\mathbb{I}_{\{U^i(\pi) \geq U^i(\tilde{\pi}^i,\pi^{-i}) - \epsilon^i,\forall \tilde{\pi}^i\}} &\mbox{w.p. } \xi^{w^i(C^i-g^i(U^i(\pi)))} \\
0 &\mbox{o.w.}
\end{array}\right.
\ee
for some $\xi \in (0,1)$ and sufficiently large $C^i \in \mathbb{R}$ such that $g^i(U^i(\pi))\leq C^i$ for all $i$ and $\pi\in\Pi$, yielding that $\xi^{w^i(C^i-g^i(U^i(\pi)))} \in (0,1]$ is a valid probability. 

\begin{remark}\label{remark:signaling}
The signaling rule \eqref{eq:signal} ensures that given joint strategy $\pi$, all agents simultaneously signal \textit{content} with the conditional probability
\be
\Pr(m^i = 1\;\forall i\mid \pi) = \left\{\begin{array}{ll}
\xi^{C-\SW(\pi)} & \mbox{if } \pi \in \Pieps\\
0 & \mbox{if } \pi \notin \Pieps 
\end{array}\right.,\label{eq:independent}
\ee
where $C := \sum_i w^i C^i$. We let agents explore their strategies uniformly such that each joint strategy gets played with equal probabilities. Then, conditioned on the event that all agents signal \textit{content}, each joint strategy $\pi$ gets played with the conditional probability
\begin{flalign}
\mu(\pi):=&\Pr(\pi\mid m^i = 1\;\forall i)
= \left\{\begin{array}{ll}
\frac{e^{\beta\cdot\SW(\pi)}}{\sum_{\tilde{\pi}\in \Pieps}e^{\beta\cdot\SW(\tilde{\pi})}} & \mbox{if } \pi \in \Pieps\\
0 & \mbox{otherwise}
\end{array}
\right.,
\end{flalign}
where $\beta:=-\ln\xi>0$ as $\xi\in(0,1)$. Let $\mu^{\mathrm{eq}}:=\mu\mid_{\Pieps}$ be a restriction of $\mu$ to $\Pieps$. Then, this yields that $\mu^{\mathrm{eq}}$ is a distribution maximizing the entropy-regularized social welfare over $\Pieps$, i.e., we have
\be
\mu^{\mathrm{eq}} = \argmax_{\tilde{\mu}\in\Delta(\Pieps)}\left\{\E_{\pi\sim \mu}[\SW(\pi)] + \frac{1}{\beta}\mathcal{H}(\mu)\right\},
\ee
where $\mathcal{H}(\mu)=-\sum_{\pi}\mu(\pi)\log(\pi)$ is the entropy function.\footnote{Let $\Delta(A)$ denote the probability simplex over a finite set $A$.} The uniqueness of the maximizer follows from the strict concavity of the objective due to the entropy regularization. Furthermore, we have
\be
\argmax_{\pi\in \Pieps} \{\mu(\pi)\} =  \argmax_{\pi\in \Pieps} \{\SW(\pi)\}.
\ee
Consequently, the optimal joint strategies $\pi_\ast$ get played with the highest probability across content-endorsed strategies. 
\end{remark}

In the following, we build our decentralized learning dynamics based on this observation.

\subsection{Explore-and-Commit-based DOEL (E\&C-DOEL)}

We first present the DOEL dynamics using the \textit{explore-and-commit} scheme. Based on Remark \ref{remark:signaling}, agents first estimate the optimal joint strategy as the most frequently played joint strategy when all agents signal \textit{content} in the uniform exploration stage over pure-stationary strategies. Then, they play according to the identified joint strategy in the remaining time. However, they face the following challenges for decentralized learning in stochastic games:
\begin{itemize}
\item For the signaling scheme, agents need $U^i(\pi)$ and $\max_{\bar{\pi}^i\in \Pi^i}\{U^i(\bar{\pi}^i,\pi^{-i})\}$ for each explored joint strategy $\pi$. However, these values represent expected discounted return accumulated over an infinite horizon. An agent cannot evaluate it 
exactly in a single stage without knowing the complete model including the underlying game and others' strategies.
\item Markov stationary pure strategies determine how agents take actions contingent on state. Therefore, only actions specific to the current state gets realized at a single stage rather than the complete strategy. They should continue to play according to the same strategy so that the complete strategy can get realized.
\item Agents cannot observe the actions or rewards of others, hindering direct inference of their utilities or strategies, and thus complicating the alignment of local (self-interested) decisions with the global objective~\eqref{eq:optimal}.
\end{itemize}

We address these challenges by dividing the exploration stage into $K$-length \textit{phases} in which agents play according to the same strategy for $K$ stages. However, pure strategies would lead to limited exploration for state-action values and best responses. For effective exploration within exploration phases, we introduce the exploration kernel transforming any pure Markov strategy $\pi^i$ to exploration-perturbed one as follows 
\be
\explore^i(\pi^i) := \{\epi^i : \epi^i(s) = (1-\varepsilon^i) \cdot \pi^i(s) + \varepsilon^i \cdot \mathrm{Unif}(A^i)\},
\ee
where $\mathrm{Unif}(A^i)$ is the uniform distribution over $A^i$ and $\varepsilon^i \in (0,1)$. For example, let agent $i$ explore $\pi_k^i\sim \mathrm{Unif}(\Pi^i)$ at phase $k$. Then, agent $i$ plays according to the perturbed strategy $\epi_k^i = \explore^i(\pi_k^i)$. 
 
At each phase $k$, agents can evaluate the value of the perturbed joint strategy $\epi_k$ and the best response by estimating value function based on the observations made within the phase. Particularly, given that agents $j\neq i$ play according to Markov stationary strategy $\epi^{-i}$, the underlying game reduces to a Markov decision process (MDP) characterized by the tuple $\langle S, A^i, \bar{r}_{\epi^{-i}}^i, \bar{p}_{\epi^{-i}}^i,\gamma \rangle$, where we define
\begin{flalign}
&\bar{r}_{\epi^{-i}}^i(s,a^i) := \E_{a^{-i}\sim\epi^{-i}(s)}[r^i(s,a^i,a^{-i})]\\
&\bar{p}_{\epi^{-i}}^i(s_+\mid s,a^i) := \E_{a^{-i}\sim\epi^{-i}(s)}[p(s_+\mid s,a^i,a^{-i})]
\end{flalign}
for all $(s,a^i,s_+)$. Given the local model $(\bar{r}^i_{\epi^{-i}},\bar{p}_{\epi^{-i}}^i)$, value function for local strategy $\pi^i$ is the \textit{unique} fixed point of the Bellman Evaluation Equation (BE):
\begin{flalign}
v^i_{\pi^i,\epi^{-i}}(s) = \E_{a^i\sim\pi^i(s)}\bigg[\bar{r}_{\epi^{-i}}^i(s,a^i)
+ \gamma\sum_{s_+\in S} \bar{p}_{\epi^{-i}}^i(s_+\mid s,a^i) \cdot v^i_{\pi^i,\epi^{-i}}(s_+)\bigg].\tag{BE}\label{eq:FPeval}
\end{flalign}
Value function for the optimal local strategy is the \textit{unique} fixed point of the Bellman Optimality Equation (BO): 
\begin{flalign}
v^i_{*,\epi^{-i}}(s) = \max_{a^i\in A^i}\bigg\{\bar{r}_{\epi^{-i}}^i(s,a^i)\nonumber
+ \gamma\sum_{s_+\in S} \bar{p}_{\epi^{-i}}^i(s_+\mid s,a^i) \cdot v^i_{*,\epi^{-i}}(s_+)\bigg\}.\tag{BO}\label{eq:FPopt}
\end{flalign}
Their uniqueness follows from the $\gamma$-contraction of the Bellman evaluation and Bellman optimality operators \citep{ref:Sutton18}. 

If the others' play is stationary, agent $i$ can estimate $v^i_{\pi_k^i,\epi_k^{-i}}$ and $v^i_{*,\epi_k^{-i}}$ in a model-based/model-free way by solving the fixed-point equations \eqref{eq:FPeval} and \eqref{eq:FPopt} based on the trajectory of state, local action and local reward realized within the phase \citep{ref:Sutton18}. Given these values and the initial state $s_0$, they can obtain
\begin{flalign}
&U^i(\pi_k^i,\epi_k^{-i}) = v^i_{\pi^i_k,\epi^{-i}_k}(s_0),\\ 
&\max_{\pi^i\in \Pi^i}\{U^i(\pi^i,\epi_k^{-i})\} = v^i_{*,\epi^{-i}_k}(s_0).
\end{flalign} 

Furthermore, at the end of each exploration phase, each agent can broadcast \textit{content} `$1$' or \textit{discontent} `$0$' based on the estimated values of the joint strategy and the best response, according to the signaling scheme described in Subsection \ref{sec:signaling}. After the exploration phases, each agent $i$ can commit to playing according to the most frequently content-endorsed local strategy $\cpi^i$. 

\begin{algorithm}[t!]
    \caption{E\&C-DOEL Dynamics for Agent $i$}
    \label{alg:main}
    \begin{algorithmic}
    \small
    \Require{$\xi\in (0,1)$, $\epsilon^i\geq 0$, $\varepsilon\in (0,1)$}
    \State \hspace{0.5em}\textbf{Initialize:} $c^i_{0}(\pi^i) = 0$ for all $\pi^i\in\Pi^i$
    \vspace{-0.4cm}
    \Statex
    \tikz[remember picture,overlay] {
        \node[rotate=90,anchor=south,yshift=5pt] at (0.2,-3.5) {\scriptsize \textbf{Exploration}};
        \draw[thick] (0,-8.1) -- (0,-0.25); 
    }
    \State \hspace{0.5em}\textbf{for} each phase $k=0,1,\ldots,\kappa-1$ \textbf{do}
        \State \hspace{2em}choose strategy $\pi_k^i \sim \mathrm{Unif}(\Pi^i)$ independently
        \State \hspace{2em}construct exploration-perturbed strategy $\epi_k^i=\explore^i(\pi_k^i)$
        \State \hspace{2em}\textbf{for} each stage $t=kK,\ldots,(k+1)K-1$ \textbf{do}
            \State \hspace{4em}observe state $s_t$ 
            \State \hspace{4em}play $a_t^i\sim\epi_{k}^i(s_t)$ \Comment{Simultaneous play}
            \State \hspace{4em}receive reward $r^i_t$
        \State \hspace{2em}\textbf{end for} \Comment{Compute and broadcast at the end of the phase}
        \State \hspace{2em}set the trajectory $h_k^i = (s_t,a_t^i,r_t^i)_{t=kK}^{(k+1)K-1}$
        \State \hspace{2em}compute $v_k^i$ = \Call{SolveBE}{$\pi_k^i,h_k^i$} 
        \State \hspace{2em}compute $v_{*,k}^i$ = \Call{SolveBO}{$h_k^i$}
        \State \hspace{2em}broadcast a single-bit feedback
        \be\nonumber
            \label{eq:message}
            m_k^i =
            \begin{cases}
                \mathbb{I}_{\{v_k^i(s_0) \geq v_{*,k}^i(s_0)-\epsilon^i\}} & \text{w.p. } \xi^{\,w^i C^i - w^i g^i(v_k^i(s_0))} \\
                0 & \text{o.w.}
            \end{cases}
         \ee
         \State \hspace{2em}set $m_k = \prod_j m_k^j$ based on the received feedback $m_k^{-i}$
        \State \hspace{2em}update $c^i_{k+1}(\pi^i) = c^i_{k}(\pi^i) + \mathbb{I}_{\{\pi^i=\pi_k^i,m_k=1\}}$ for all $\pi^i$ 
    \State \hspace{0.5em}\textbf{end for}
    \vspace{0.1cm}
    \State \hspace{0.5em}identify $\cpi^i \in \argmax_{\pi^i \in \Pi^i} \{c_\kappa^i(\pi^i)\}$ \Comment{Any tie-breaking rule}
    \vspace{-0.3cm}
    \Statex
    \tikz[remember picture,overlay] {
        \node[rotate=90,anchor=south,yshift=5pt] at (0.2,-.75) {\scriptsize \textbf{Exploitation}};
        \draw[thick] (0,-2.1) -- (0,0.33); 
    }
    \Statex \hspace{0.5em}\textbf{for} each stage $t \geq \kappa K$ \textbf{do}
         \State \hspace{2em}observe state $s_t$ 
         \State \hspace{2em}play $a_t^i\sim\cpi^i(s_t)$ \Comment{Simultaneous play}
         \State \hspace{2em}receive reward $r^i_t$
    \State \hspace{0.5em}\textbf{end for}
    \end{algorithmic}
\end{algorithm}

Algorithm \ref{alg:main} tabulates the complete description of the E\&C-DOEL dynamics for agent $i$ with the (rather standard) sub-procedures:
\begin{itemize}
\item \Call{SolveBE}{$\cdot$} solves \eqref{eq:FPeval}, e.g.,  by iterating the Bellman evaluation operator,
\item \Call{SolveBO}{$\cdot$} solves \eqref{eq:FPopt}, e.g., by iterating the Bellman optimality operator,
\end{itemize}
given the trajectory $h_k^i=(s_t,a_t^i,r_t^i)_{t=kK}^{(k+1)K-1}$ realized. 

\begin{remark}
For the sub-procedures  \Call{SolveBE}{$\cdot$} and \Call{SolveBO}{$\cdot$}, agents have various alternatives from model-free methods, such as Q-learning and Monte Carlo sampling, to model-based dynamic programming methods, such as value and policy iterations, using the model estimated based on the trajectory sampled \citep{ref:Sutton18}. We present a framework in which agents locally choose their (possibly different) methods. If they adopt iterative methods, these iterations can evolve over a timescale different from the timescale of the underlying stochastic game with fixed or threshold-based termination.

These procedures compute $v_k^i$ and $v_{*,k}^i$ for signaling \textit{content}/\textit{discontent}. Notably, if agents store the trajectory of the explored local strategies, they could have shared feedback signals $\{m_k^i\}_{k=0}^{\kappa-1}\in \{0,1\}^{\kappa}$ collectively at the end of the exploration at stage $t=\kappa K$ once, e.g., via pre-determined encoding and decoding schemes that can ensure efficient and reliable transmission. Therefore, there is no strict time-constraint on these procedures to complete their computations in-between the stages $(k+1)K-1$ (the end of phase $k$) and $kK$ (the start of phase $k+1$).
\end{remark}

\begin{algorithm}[t!]
    \caption{Online-DOEL Dynamics for Agent $i$}
    \label{alg:online}
    \begin{algorithmic}
    \small
    \Require{$\xi\in (0,1)$,$\epsilon^i\geq 0$, $\varepsilon\in (0,1)$, $\rho_k^i\in (0,1)$}
    \State \textbf{Initialize:} $c^i_{0}(\pi^i) = 0$ for all $\pi^i\in\Pi^i$
    \For{each phase $k=0,1,\ldots$}
    \State broadcast $\bar{m}_k^i=1$ w.p. $\rho_k^i$ and $\bar{m}_k^i=0$ o.w. independently
    \If{$\bar{m}_k^j = 1$ for some $j$} 
    \State choose strategy $\pi_k^i \sim \mathrm{Unif}(\Pi^i)$ independently
    \State construct exploration-perturbed strategy $\epi_k^i=\explore^i(\pi_k^i)$
    \For{each stage $t=kK,\ldots,(k+1)K-1$}
            \State observe state $s_t$ 
            \State play $a_t^i\sim\epi_{k}^i(s_t)$ \Comment{Simultaneous play}
            \State receive reward $r^i_t$
     \EndFor  \Comment{Compute and broadcast at the end of the phase}
     \State set the trajectory $h_k^i = (s_t,a_t^i,r_t^i)_{t=kK}^{(k+1)K-1}$
        \State compute $v_k^i$ = \Call{SolveBE}{$\pi_k^i,h_k^i$} 
        \State compute $v_{*,k}^i$ = \Call{SolveBO}{$h_k^i$}
     \State broadcast \textit{content/discontent}
        \be\nonumber
            m_k^i =
            \begin{cases}
                \mathbb{I}_{\{v_k^i(s_0) \geq v_{*,k}^i(s_0)-\epsilon^i\}} & \text{w.p. } \xi^{\,w^i C^i - w^i g^i(v_k^i(s_0))} \\
                0 & \text{o.w.}
            \end{cases}
         \ee
            \State set $m_k = \prod_j m_k^j$ based on the received feedback $m_k^{-i}$
        \State update $c^i_{k+1}(\pi^i) = c^i_{k}(\pi^i) + \mathbb{I}_{\{\pi^i=\pi_k^i,m_k=1\}}$ for all $\pi^i$
    \Else\Comment{All exploit if all signal \textit{exploit}}
    \State identify $\cpi^i_k \in \argmax_{\pi^i} c^i_k(\pi^i)$ \Comment{Any tie-breaking rule}
    \State update $c^i_{k+1}(\pi^i) = c^i_{k}(\pi^i)$ for all $\pi^i$
    \For{each stage $t=kK,\ldots,(k+1)K-1$}
            \State observe state $s_t$ 
            \State play $a_t^i\sim\cpi_{k}^i(s_t)$ \Comment{Simultaneous play}
            \State receive reward $r^i_t$
     \EndFor 
    \EndIf
    \EndFor
    \end{algorithmic}
\end{algorithm}

\subsection{Online DOEL} 
Next, we present a more practical variant for the E\&C-DOEL dynamics in which exploration and exploitation are intertwined. Across phases, each agent $i$ can independently explore with certain probability $\rho_k^i\in(0,1)$ and exploit otherwise. At the beginning of each phase, they broadcast a single-bit \textit{explore} `$1$' or \textit{exploit} `$0$' to the others to show their intention within the phase. To ensure joint exploration and exploitation, they all collectively explore if at least one agent signals \textit{explore}. Notably, the exploration probability $\rho_k^i$ can be agent-specific and time-dependent.  

Apart from the signaling scheme for the coordination on joint exploration and exploitation, E\&C-DOEL and its online variant Online-DOEL follow identical steps for exploration and exploitation. Algorithm \ref{alg:online} tabulates the complete description of the Online-DOEL dynamics for agent $i$.

\begin{remark}
The pure stationary strategy space $\Pi^i$ has cardinality $|A^i|^{|S|}$, which is finite yet exponential in $|S|$. In Algorithms \ref{alg:main} and \ref{alg:online}, uniform sampling from $\Pi^i$ acts as a \textit{conceptual mechanism} to guarantee full-support exploration over all Markov strategies, analogous to uniform action sampling in finite games. Importantly, agents only sample a single strategy per phase and update counters for strategies actually realized. Thus, the memory and computational load scale with the number of phases, not with $|\Pi^i|$. Practical implementations may further restrict the strategy class (e.g., via parametric strategies) for sub-optimal solutions without altering the structure of the DOEL framework.
\end{remark}

\section{Main Results}\label{sec:results}
In this section, we present the main performance guarantees for Algorithms \ref{alg:main} and \ref{alg:online}. In these guarantees, pure Markov stationary strategies $\pi^i:S\rightarrow A^i$ play an important role. Particularly, the strategy set $\Pi^i = (A^i)^S$ is \textit{finite} and, therefore, we can equivalently represent the underlying game $\langle S,\{A^i,r^i\}_{i\in N},p,\gamma\rangle$ as a \textit{normal-form} game $\langle \Pi^i,U^i\rangle_{i\in N}$. However, in the DOEL dynamics, agents are not actually playing pure strategies due to the $\varepsilon$-exploration. Therefore, we define $\eU^i(\pi^i,\pi^{-i}) := U^i(\pi^i,\explore^{-i}(\pi^{-i}))$ for all $i$ and $\pi$, where $\explore^{-i}(\pi^{-i}) = (\explore^j(\pi^j))_{j\neq i}$, as the \textit{effective utility} under the exploration-perturbed execution. Correspondingly, we define the effective social welfare as
\be\label{eq:eSW}
\eSW(\pi) := \sum_{i=1}^n w^i \cdot g^i(\eU^i(\pi)). 
\ee
Furthermore, the effective $\epsilon$-equilibrium set is given by
\be\nonumber
\ePieps := \{\pi\in \Pi: \eU^i(\pi^i,\pi^{-i})\geq \max_{\tilde{\pi}^i\in \Pi^i} \{\eU^i(\tilde{\pi}^i,\pi^{-i})\} - \epsilon^i,\;\forall i\}
\ee
rather than \eqref{eq:bestset}.

 Our analysis focusses on the effective normal-form game $\langle \Pi^i,\eU^i\rangle_{i\in N}$ and rely on the following assumptions. 

\begin{assumption}\label{assm:SW}
For each agent $i$,
\begin{itemize}
\item \label{assm:SWi} The parameter $C^i$ is sufficiently large such that $C^i-g^i(\eU^i(\pi)) \geq 0$ for all $\pi\in \Pi$ and, therefore, $C=\sum_i w^i C^i \geq \eSW(\pi)$ for all $\pi$,
\item \label{assm:SWii} The transformation $g^i$ is $L_g$-Lipschitz continuous.
\end{itemize}
\end{assumption}

Assumption \ref{assm:SW}-i ensures that the signaling probabilities in the stochastic signaling rule \eqref{eq:signal} are valid probabilities. Assumption \ref{assm:SW}-ii restrains the impact of the errors induced by the sub-procedures \Call{SolveBE}{$\cdot$} and \Call{SolveBO}{$\cdot$}, and the exploration-perturbed execution on the social welfare.

\begin{assumption}\label{assm:optimal}
Effective equilibrium strategies satisfy:
\begin{itemize}
\item The effective equilibrium set $\ePieps$ is non-empty and there exists a \textit{unique} effective $\epsilon$-equilibrium $\epi^*$ maximizing the effective social welfare:
\be\label{eq:separate}
\eSW(\epi^*) > \eSW(\pi)\quad\forall \pi \in \ePieps\mbox{ and } \pi\neq \epi^*.
\ee
\item For some $\delta > 0$, we have
\be
0< \xi < (|\ePieps|+\delta)^{-1/\Xi_o},
\ee
where the gap $\Xi_o$ between the best and second best(s) is given by
\be\label{eq:DeltaSW}
\Xi_o := \eSW(\epi^*) - \max_{\pi\in \ePieps\setminus \epi^*} \{\eSW(\pi)\}
\ee
and $\Xi_o$ is positive due to the unique solution in \eqref{eq:separate}.
\item There exists $\eta_o>0$ such that for any joint strategy $\pi$, 
\be
\eU^i(\pi) \geq \max_{\tilde{\pi}^i\in \Pi^i}\eU^i(\tilde{\pi}^i,\pi^{-i}) - \epsilon^i + \eta_o 
\ee
for all $i$ if $\pi \in \ePieps$ and
\be
\eU^i(\pi) < \max_{\tilde{\pi}^i\in \Pi^i}\eU^i(\tilde{\pi}^i,\pi^{-i}) - \epsilon^i - \eta_o 
\ee 
for some $i$ if $\pi\notin\ePieps$.
\end{itemize}
\end{assumption}

Assumptions \ref{assm:optimal}-i and \ref{assm:optimal}-ii play important roles in ensuring the existence and uniqueness of the optimal joint strategy and its sufficient separation from other strategies so that agents can learn it based on local information only in a decentralized way since local strategies may not be transferable across multiple optimal joint strategies. Correspondingly, we can relax the uniqueness condition if agents can observe the opponent actions to keep track of the content-endorsed joint strategies and coordinate with the others to play according to the same optimal strategy. On the other hand, Assumption \ref{assm:optimal}-iii is a non-degeneracy assumption that excludes knife-edge cases where an $\epsilon$-equilibrium lies exactly on the boundary of $\eqref{eq:best}$ to address the estimation errors in the value function estimates. For (finite) pure strategies, such degeneracies can be removed by an arbitrarily small perturbation of payoffs. 

\begin{assumption*}{Per-phase Accuracy}\label{assm:accuracy}
The phase length $K$ is sufficiently long such that agent $i$'s value estimates satisfy
\begin{subequations}\label{eq:accuracy}
\begin{flalign}
&|v_k^i(s_0)-\eU^i(\pi_k)| \leq \eta \\
&|v_{*,k}^i(s_0) - \max_{\tilde{\pi}^i\in \Pi^i}\{\eU^i(\tilde{\pi}^i,\pi_k^{-i})\}|\leq \eta
\end{flalign}
\end{subequations}
with probability at least $1-\rho$ for constants $c_1,c_2 > 0$ and certain $\eta,\rho>0$ satisfying $2\eta < \eta_o$, where $\eta_o$ is as described in Assumption \ref{assm:optimal}-iii, and
\be\label{eq:zeta}
L \eta + \rho < \frac{\delta}{|\Pi||\ePieps|}\cdot \xi^{C-\max_{\pi\neq \epi_*}\eSW(\pi)} =:\zeta 
\ee
where $L:= \frac{1}{2}|\log \xi| L_g\sum_i w^i$.
\end{assumption*}

Achieving $\eta$-accuracy from a $K$-length trajectory implicitly requires sufficient state--action coverage and mixing/ergodicity on the reachable component—otherwise some values are not identifiable from a single trajectory—while persistent exploration (e.g., a uniform lower bound on action probabilities) typically ensures these conditions on the reachable class. Hence Assumption~\ref{assm:accuracy} can be satisfied by choosing $K$ large enough (as a function of $\eta,\rho,\gamma$ and a mixing/coverage term) and instantiating \textsc{SolveBE}/\textsc{SolveBO} with standard model-free SA/$Q$-learning routines under Markovian noise or with model-based plug-in estimation of $(\bar r,\bar p)$ followed by planning in the empirical MDP, e.g., see \citep{ref:Qu20,ref:Agarwal20}. 

For the E\&C-DOEL dynamics, the following theorem shows that expected regret is bounded by $O(\log T)$ for $O(\log T)$ exploration lengths. Here, the expectation is taken with respect to the randomness induced by the randomized exploration and signaling.  

\begin{theorem}\label{thm:main}
Consider an $n$-agent stochastic game $\langle S, \{A^i,r^i\}_{i\in N}, p, \gamma\rangle$. Suppose that each agent $i$ follows Algorithm \ref{alg:main} and Assumptions \ref{assm:SW}, \ref{assm:optimal}, and \ref{assm:accuracy} hold. If the exploration length $\kappa$ satisfies
$\kappa = \log(8|\Pi|T)/(2\tilde{\zeta}^2) \in O(\log T)$,
where $\tilde{\zeta}>0$ is a constant as described in Proposition \ref{prop:probability}, then the expected regret is bounded by
\be
\E[R_T] \leq C_1 + C_2 \log T + \overline{C}\cdot \varepsilon\cdot T
\ee
for constants $C_1,C_2>0$ as described in \eqref{eq:CC} and $\overline{C}>0$ as described in \eqref{eq:oC}.
\end{theorem}

For the Online-DOEL dynamics, the following theorem shows that expected regret is also bounded by $O(\log T)$ for exploration probabilities decaying sufficiently slowly. 

\begin{theorem}\label{thm:online}
Consider an $n$-agent stochastic game $\langle S, \{A^i,r^i\}_{i\in N}, p, \gamma\rangle$. Suppose that each agent $i$ follows Algorithm \ref{alg:online} and Assumptions \ref{assm:SW}, \ref{assm:optimal}, and \ref{assm:accuracy} hold. If the exploration probabilities $\rho_k^i=\min\{1,c/(nk)\}$ for each $i$ with $c >n/ (1-e^{-2\tilde{\zeta}^2})$ and $\tilde{\zeta}$ is as described in Proposition \ref{prop:probability}, then the expected regret is bounded by
\be
\E[R_T] \leq C_1' + C_2' \log T + \overline{C}\cdot \varepsilon\cdot T
\ee
for constants $C'_1,C'_2>0$ as described in \eqref{eq:CCC} and $\overline{C}>0$ as described in \eqref{eq:oC}.
\end{theorem}

The proofs of Theorems \ref{thm:main} and \ref{thm:online} are deferred to Section \ref{sec:proofs}. 

\begin{remark}
In Algorithms \ref{alg:main} and \ref{alg:online}, agents use local parameters and update rules, except a common parameter $\xi$ used in the signaling mechanism. Consider that each agent $i$ has possibly different $\xi^i$. Then, given a joint strategy $\pi\in \Pieps$, all would have signaled \textit{content} with the conditional probability
\be\label{eq:newindependent}
\prod_{i=1}^n (\xi^i)^{w^i C^i - w^i g^i(U^i(\pi))}
\ee
rather than \eqref{eq:independent}. Let $\overline{\xi}:=\max_i \xi^i$. Then, we can write \eqref{eq:newindependent} as
\be
\prod_{i=1}^n \overline{\xi}^{\overline{w}^i C^i - \overline{w}^i g^i(U^i(\pi))} = \overline{\xi}^{\sum_i \overline{w}^i C^i - \sum_{i} \overline{w}^i g^i(U^i(\pi))}
\ee
with the modified weights $\overline{w}^i := w^i \frac{\log \xi^i}{\log \overline{\xi}} > 0$.
Therefore, Theorems \ref{thm:main} and \ref{thm:online} yield that Algorithms \ref{alg:main} and \ref{alg:online} would have learned the effective equilibrium maximizing the modified welfare $\sum_{i} \overline{w}^i g^i(U^i(\pi))$. Furthermore, the difference between the original and modified welfare is small if agents have similar $\xi^i$'s and agents can increase (or decrease) their share in the modified welfare with smaller (or larger) $\xi^i$.
\end{remark}

\section{Proofs of Theorems \ref{thm:main} and \ref{thm:online}}\label{sec:proofs}

The proofs for Theorem \ref{thm:main} and \ref{thm:online} follow a common template: We rewrite regret in terms of exploration phases,  characterize the committed strategies, i.e., the most frequent content-endorsed strategies, in comparison to the optimal (effective) equilibrium $\epi_*$, and finally determine $\kappa$ (or $\{\rho_k\}$) to make the deviation probability $O(1/T)$ yielding logarithmic regret. 

Both Algorithms \ref{alg:main} and \ref{alg:online} divide the horizon into $K$-length phases. Without loss of generality, let $T$ be an integer multiple of $K$ and $\tau:= T/K$. Since agents play the same strategy within phases, we can write the regret \eqref{eq:regret} as
\be\label{eq:regretphased}
R_T = \tau K \cdot \SW^* - K \sum_{k=1}^\tau \SW(\tilde{\pi}_k),
\ee
where $\SW^*:= \max_{\pi\in\Pieps}\SW(\pi)$ and $\tilde{\pi}_k$ is the executed strategy. For example, in phase $k$, $\tilde{\pi}_k = \epi_k=(\explore^i(\pi_k^i))_{i\in N}$ is the exploration-perturbed version of the explored strategy $\pi_k\sim \mathrm{Unif}(\Pi)$ in exploration phases while $\tilde{\pi}_k = \cpi_k$ is the committed strategy in the exploitation phases.  

Denote the index set of exploration phases by $\taue\subset \{1,\ldots,\tau\}$. Then, we can write the expected regret as
\begin{equation}
\E[R_T] = K \cdot \E\bigg[\sum_{k\in \taue} (\SW^* - \SW(\epi_k))\bigg] + K\cdot\E\bigg[\sum_{k\notin \taue} (\SW^* - \SW(\cpi_k))\bigg].\label{eq:regretphased2}
\end{equation}
We define $\Xi := \SW^* - \min_{\pi}\{\SW(\pi)\}$ and $\Xi_\varepsilon := \SW^* - \SW(\epi_*)$. Based on $\Xi, \Xi_\varepsilon>0$ and the total probability theorem, we can bound \eqref{eq:regretphased2} from above by 
\begin{equation}
\E[R_T] \leq K \Xi \cdot \E[|\taue|] + T\Xi_\varepsilon + K \Xi \cdot \E\bigg[\sum_{k\notin \taue}\Pr(\cpi_k\neq\epi_*\mid \taue)\bigg],\label{eq:bound}
\end{equation}
where $\epi_*$ is the (unique) optimal effective equilibrium, as described in Assumption \ref{assm:SWi}. The following lemma characterizes $\Xi_\varepsilon$ in terms of $\varepsilon$ and its proof is moved to Appendix \ref{app:SW}

\begin{lemma}\label{lem:SW}
Under Assumption \ref{assm:SW}-ii, we have $0<\Xi_\varepsilon\leq \overline{C} \varepsilon$, where $\overline{C} >0$ is as described in \eqref{eq:oC}.
\end{lemma}

In Algorithm \ref{alg:main}, the index set $\taue$ is deterministic and we have $\taue = \{1,\ldots,\kappa\}$. However, $\taue$ is random in Algorithm \ref{alg:online}. Independent and identical exploration across exploration phases yield that the probability of mismatch between the committed strategy and the optimal effective equilibrium is independent of the exact indices of the exploration phases given the number of exploration phases until phase $k$. Let $\taue=\{\tau_1,\tau_2,\ldots,\}$, where $\tau_\ell$ denotes the phase index for the $\ell$th exploration. Then, we have
\be
\Pr(\cpi_k\neq\epi_*\mid \taue=\{\tau_1,\tau_2,\ldots\}) = \Pr(\cpi_{\tau_\ell}\neq\epi_*)
\ee 
for all $k\in (\tau_\ell,\tau_{\ell+1})$. The following proposition provides an upper bound on this probability.

\begin{proposition}\label{prop:probability}
Under Assumptions \ref{assm:SW}, \ref{assm:optimal}, and \ref{assm:accuracy}, we have
\be\label{eq:probability}
\Pr(\cpi_{\tau_\kappa}\neq\epi_*) \leq 8|\Pi|\exp(-2 \tilde{\zeta}^2\cdot \kappa ),
\ee
where $\tilde{\zeta}:= \frac{1}{2}(\zeta-L\eta - \rho)>0$, after $\kappa$ exploration phases.
\end{proposition}

\begin{myproof}
Denote the empirical average of the joint strategy $\pi\in \Pi$ and joint content-endorsement feedback $m\in\{0,1\}$ after $\ell$ exploration phases by
\begin{flalign}\label{eq:wtheta}
\wtheta_\ell(\pi,m) := \frac{1}{\ell}\sum_{l=1}^\ell \mathbb{I}_{\{\pi_{\tau_l} = \pi,m_{\tau_l}=m\}}.
\end{flalign}
Then, the empirical average of the local strategy and joint feedback $m$ is given by
\be\label{eq:marginalization}
\wtheta_\ell^i(\pi^i,m) := \frac{1}{\ell}\sum_{l=1}^\ell \mathbb{I}_{\{\pi_{\tau_l}^i = \pi^i,m_{\tau_l}=m\}} = \sum_{\pi^{-i}} \wtheta_\ell(\pi^i,\pi^{-i},m)
\ee
and the committed strategy is given by
\be
\cpi_{k}^i \in \argmax_{\pi^i\in \Pi^i} \{\wtheta_\ell^i(\pi^i,1)\}\quad\forall k\in [\tau_\ell,\tau_{\ell+1})
\ee
by the definitions of the counter $c_k^i(\pi)$ and the committed strategy $\cpi_k^i$ in Algorithms \ref{alg:main} and \ref{alg:online}.
Therefore, the frequency that the pairs $(\pi,m=1)$ for $\pi\in \Pi$ get realized in exploration phases determines the committed strategy. However, the per-phase accuracy of the value estimates $v_k^i,v_{*,k}^i$ pose a challenge. 

As a benchmark, we first focus on the \textit{ideal} case where $v_k^i(s_0) = \eU^i(\pi_k)$ and $v_{*,k}^i(s_0) = \max_{\pi^i}\eU^i(\pi^i,\pi_k^{-i})$. In the ideal case, at each exploration phase, the pair $(\pi,m)\in \Pi\times \{0,1\}=:Z$ would have been drawn from the distribution 
\be\label{eq:joint}
\mathcal{P}(\pi,m) := \left\{\begin{array}{ll}
\frac{1}{|\Pi|}\xi^{C - \eSW(\pi)} & \mbox{if } \pi \in \ePieps, m=1\\
0 & \mbox{if } \pi \notin \ePieps, m=1\\
\frac{1}{|\Pi|}(1-\xi^{C - \eSW(\pi)}) & \mbox{if } \pi \in \ePieps, m=0\\
\frac{1}{|\Pi|} & \mbox{if } \pi \notin \ePieps, m=0
\end{array}\right.
\ee
as discussed in Remark \ref{remark:signaling}. However, different from the conditional probability \eqref{eq:independent}, the joint probability \eqref{eq:joint} has $1/|\Pi|$ scaling factor due to uniform exploration $\pi^i\sim \mathrm{Unif}(\Pi^i)$ for all $i$ and has the effective social welfare $\eSW$, as described in \eqref{eq:eSW}, rather than $\SW$ due to the in-phase exploration. Observe that $\epi_* = \argmax_{\pi} \mathcal{P}(\pi,1)$.

On the other hand, in the \textit{actual} scenario, the per-phase accuracy errors
\begin{subequations} 
\begin{flalign}
&e_k^i := v_k^i(s_0) - \eU^i(\pi_k),\\  
&e_{*,k}^i := v_{*,k}^i(s_0) - \max_{\pi^i}\eU^i(\pi^i,\pi_k^{-i})
\end{flalign}
\end{subequations}
are random due to their dependence on the trajectory of states, actions and rewards within the associated phase. To incorporate this randomness, we introduce a kernel $\mathcal{V}$ such that $\nu(\pi_k):=(v_k^i(s_0;\pi_k),v_{*,k}^i(s_0;\pi_k))\sim \mathcal{V}(\pi_k)$. Here, we explicitly show the implicit dependence of value estimates on the strategy $\pi_k$. Then, given $\nu(\pi)\sim \mathcal{V}(\pi)$ for each $\pi$, the pair $(\pi,m)\in Z$ gets drawn from the distribution
\be\label{eq:jointv}
\mathcal{P}_\nu(\pi,m) := \left\{\begin{array}{ll}
\frac{1}{|\Pi|}\xi^{C - \eSWv(\pi)} & \mbox{if } \pi \in \ePiepsv, m=1\\
0 & \mbox{if } \pi \notin \ePiepsv, m=1\\
\frac{1}{|\Pi|}(1-\xi^{C - \eSWv(\pi)}) & \mbox{if } \pi \in \ePiepsv, m=0\\
\frac{1}{|\Pi|} & \mbox{if } \pi \notin \ePiepsv, m=0
\end{array}\right.
\ee
where
\begin{subequations}
\begin{align}
&\eSWv(\pi) := \sum_{i=1}^n w^i\cdot g^i(v^i(s_0;\pi)),\\
&\ePiepsv := \{\pi\in \Pi: v^i(s_0;\pi) \geq v_*^i(s_0;\pi) - \epsilon^i\;\forall i\}.
\end{align}
\end{subequations}

Similar to \eqref{eq:wtheta} in the actual scenario, we denote the empirical average of the joint strategy and the joint feedback in the ideal scenario by
\be
\wtheta^\ideal_\kappa (z) := \frac{1}{\kappa}\sum_{l=1}^\kappa \mathbb{I}_{\{z=z_l^\ideal\}}
\ee
for all $z\in Z$. If we consider the indicator function for $\{z=z_l^\ideal\}$, the Hoeffding bound for Bernoulli random variables yields that
\be\label{eq:bound1}
\Pr\big(|\wtheta^\ideal_\kappa(z) - \mathcal{P}(z)| \geq c\big) \leq 2 \exp(-2\kappa c^2) \quad\forall c > 0.
\ee

Within ideal or actual scenarios, the samples are independent across exploration phases due to independent randomization of explored strategies and the feedback signals. In the following lemma, we couple the samples for the ideal and actual cases to bound the distance between the empirical averages $\wtheta^\ideal_\kappa$ and $\wtheta_\kappa$. Its proof is moved to Appendix \ref{app:distance}.

\begin{lemma}\label{lem:distance}
Under Assumptions \ref{assm:SW}, \ref{assm:optimal}, and \ref{assm:accuracy}, we have
\be\label{eq:bound2}
\Pr\big(|\wtheta_\kappa(z) - \wtheta_\kappa^\ideal(z)| \geq c + L \eta + \rho\big) \leq 2 \exp(-2\kappa c^2)
\ee
for all $c>0$.
\end{lemma}

We can quantify the distance between the actual empirical distribution and $\mathcal{P}$ of the ideal scenario as 
\begin{flalign}
\Pr\big(|\wtheta_\kappa(z) - \mathcal{P}(z)| < 2c + L \eta + \rho\big) 
\geq & \; \Pr\big(|\wtheta_\kappa(z) - \wtheta_\kappa^\ideal(z)| < c + L\eta + \rho\big) \nonumber 
\\&+ \Pr\big(|\wtheta_\kappa^\ideal(z) - \mathcal{P}(z)| < c\big) - 1\\
\geq & \; 1 - 4 \exp(-2\kappa c^2)\quad\forall c>0,
\end{flalign}
where the first inequality follows from the triangle inequality and $\Pr(A\cap B) \geq \Pr(A)+\Pr(B)-1$ while the second one is due to \eqref{eq:bound1} and \eqref{eq:bound2}. Correspondingly, we have
\be\label{eq:bound3}
\Pr\big(|\wtheta_\kappa(z) - \mathcal{P}(z)| \geq 2c + L\eta + \rho\big) \leq 4 \exp(-2\kappa c^2)
\ee
for all $c>0$. This inequality characterizes a probabilistic bound on the closeness of the (actual) empirical average $\wtheta$ to $\mathcal{P}$. The following lemma shows that the local committed strategies $\cpi_\kappa^i$ are the optimal (effective) equilibrium $\epi_*$ if $\wtheta$ is sufficiently close to $\mathcal{P}$.

\begin{lemma}\label{lem:separation}
Under Assumptions \ref{assm:optimal}-i and \ref{assm:optimal}-ii, if $|\wtheta_\kappa(z) - \mathcal{P}(z)| < \zeta$ for all $z$, then there exist unique committed local strategies $\cpi^i_\kappa$ for each $i$ such that
$\wtheta_\kappa^i(\cpi^i_\kappa,1) > \wtheta_\kappa^i(\pi^i,1)$ for all $\pi^i\neq \cpi_{\kappa}^i$ and the 
joint committed strategy $\cpi_\kappa = (\cpi^i_\kappa)_{i\in N}$ is equal to the unique (effective) solution $\epi_*$, i.e., $\cpi_\kappa = \epi_*$.
\end{lemma}

By \eqref{eq:bound3}, we have
\be
\Pr\big(|\wtheta_\kappa(z) - \mathcal{P}(z)| \geq \zeta \big) \leq 4\exp(-2\kappa \tilde{\zeta}^2)
\ee
where $\zeta >0$ is as described in \eqref{eq:zeta}, $\tilde{\zeta}>0$ is as described in Proposition \ref{prop:probability} and it is positive by Assumption \ref{assm:accuracy}. Then, using the union bound, we obtain
\begin{flalign}
\Pr\big(|\wtheta_\kappa(z) - \mathcal{P}(z)| \geq \zeta \mbox{ for some $z\in Z$}\big) \leq 8|\Pi|\exp(-2\kappa \tilde{\zeta}^2)\nonumber
\end{flalign}
Correspondingly, we have
\begin{flalign}
\Pr\big(|\wtheta_\kappa(z) - \mathcal{P}(z)| \leq \zeta \;\forall z\in Z\big) \geq 1 - 8|\Pi|\exp(-2\kappa \tilde{\zeta}^2).\nonumber
\end{flalign}
Then, Lemma \ref{lem:separation} yields \eqref{eq:probability}, which completes the proof of Proposition \ref{prop:probability}.
\end{myproof}

\subsection{Final Step for Theorem \ref{thm:main}} 

Recall that for Algorithm \ref{alg:main}, we have the deterministic exploration set $\taue = \{1,\ldots,\kappa\}$. Then, the bound \eqref{eq:bound}, Lemma \ref{lem:SW}, and Proposition \ref{prop:probability} yield that 
\begin{equation*}
\E[R_T] \leq T\overline{C}\cdot \varepsilon + K \Xi \cdot \kappa + T \Xi\cdot 8 |\Pi|  \exp(-2\tilde{\zeta}^2\cdot \kappa). 
\end{equation*}
Choose $\kappa$ so that mis-commit probability is $O(1/T)$, i.e.,
\be
8 |\Pi|  \exp(-2\tilde{\zeta}^2\cdot \kappa) \leq \frac{1}{T}\quad\Rightarrow\quad \kappa \geq \frac{1}{2\tilde{\zeta}^2}\log(8|\Pi|T).
\ee
For such $\kappa$, we have
$\E[R_T] \leq C_1 + C_2 \log T + \overline{C}\cdot \varepsilon\cdot T,$
where
\be\label{eq:CC}
C_1 := \Xi\left(1 + \frac{K}{2\tilde{\zeta}^2}\log(8|\Pi|)\right)\quad\mbox{and}\quad C_2 := \frac{K\Xi}{2\tilde{\zeta}^2}.
\ee
This completes the proof of Theorem \ref{thm:main}.

\subsection{Final Step for Theorem \ref{thm:online}} 
For Algorithm \ref{alg:online}, we can write \eqref{eq:bound} as
\begin{equation*}
\E[R_T] \leq T\overline{C}\cdot \varepsilon + K \Xi \cdot \E[L_{\tau}] + K\Xi\sum_{k=1}^\tau\E[\Pr(\cpi_k\neq \epi_*| L_k)],
\end{equation*}
where $L_k$ denotes the (random) number of exploration phases up to (and including) phase $k$ and, therefore, $L_\tau = |\taue|$. Agents explore with probability $\overline{\rho}_k:= 1 - \prod_{i=1}^n(1-\rho_k^i)$ at stage $k$. Then, the Bernoulli exploration indicators yield that
\be\label{eq:Ltau}
\E[L_{\tau}] = \sum_{l=1}^\tau \overline{\rho}_l.
\ee
By Proposition \ref{prop:probability}, we have
\be
\E[\Pr(\cpi_k\neq \epi_*| L_k)] \leq 8|\Pi| \E\Big[e^{-2\tilde{\zeta}^2\cdot L_k}\Big]
\ee
and
\begin{equation}
\E\Big[e^{-2\tilde{\zeta}^2\cdot L_k}\Big] = \prod_{l=1}^k \Big((1-\overline{\rho}_l) + \overline{\rho}_l e^{-2\tilde{\zeta}^2}\Big) \leq \exp\Big(-(1-e^{-2\tilde{\zeta}^2})\sum_{l=1}^k \overline{\rho}_l\Big)\label{eq:expbound}
\end{equation} 
since the exploration indicators across phases are independent and $\log(1-x) \leq -x$ for $x\in (0,1)$. 
The following lemma formulates upper and lower bounds on $\sum_{l=1}^k \overline{\rho}_l$ in \eqref{eq:Ltau} and \eqref{eq:expbound}, and its proof is moved to Appendix \ref{app:sum}.

\begin{lemma}\label{lem:sum}
Given that $\rho_k^i = \min \{1,c/(nk)\}$, we have $\sum_{l=1}^k \overline{\rho}_k = k$ for $k\leq c/n$ and
\be
\overline{c} + c \log k \geq \sum_{l=1}^k \overline{\rho}_k \geq  \underline{c} + \frac{c}{n}\log k
\ee
for $k > c/n$, where $\overline{c} := \floor{c/n} - c\log \floor{c/n}$ and $\underline{c}:=\floor{c/n} - \frac{c/n}{\floor{c/n}} - \frac{c}{n}\log \floor{c/n}$. 
\end{lemma}

Based on Lemma \ref{lem:sum}, we can write \eqref{eq:Ltau} as
\be
\E[L_{\tau}]  \leq \overline{c} + c \log \tau
\ee
for $\tau > c/n$, and bound \eqref{eq:expbound} from above by
\be
\E\Big[e^{-2\tilde{\zeta}^2\cdot L_k}\Big] \leq \left\{\begin{array}{ll}
\beta k^{-\alpha}&\mbox{if } k > c/n\\
\omega & \mbox{if } k < c/n
\end{array}\right.,
\ee
where $\alpha:= c(1-e^{-2\tilde{\zeta}^2})/n$, $\beta:= \exp(-\underline{c}(1-e^{-2\tilde{\zeta}^2}))$ and $\omega:= \exp(-(1-e^{-2\tilde{\zeta}^2}))$. Note that $c>n/(1-e^{-2\tilde{\zeta}^2})$ in Theorem \ref{thm:online}. Therefore, we have
\begin{flalign}
\sum_{k=1}^\tau \E\Big[e^{-2\tilde{\zeta}^2\cdot L_k}\Big] &\leq \frac{c}{n}\omega + \sum_{k=1}^\tau \beta k^{-\alpha} \leq \beta\left(1 + \int_1^\infty x^{-\alpha}dx\right) = \frac{c}{n}\omega + \frac{\alpha\beta}{\alpha-1}.
\end{flalign}
Combining them together, we obtain $\E[R_T] \leq C_1' + C_2' \log T + \overline{C}\cdot \varepsilon\cdot T$, where 
\begin{subequations}\label{eq:CCC}
\begin{flalign}
&C_1' := K\Xi \left(\overline{c} - c \log K + 8|\Pi| \left(\frac{c}{n}\omega + \frac{\alpha\beta}{\alpha-1}\right)\right)\\
&C_2' := K\Xi c.
\end{flalign}
\end{subequations}
This completes the proof of Theorem \ref{thm:online}.

\begin{figure*}[!t]
  \centering
  \begin{subfigure}{0.48\textwidth}
    \includegraphics[width=\linewidth]{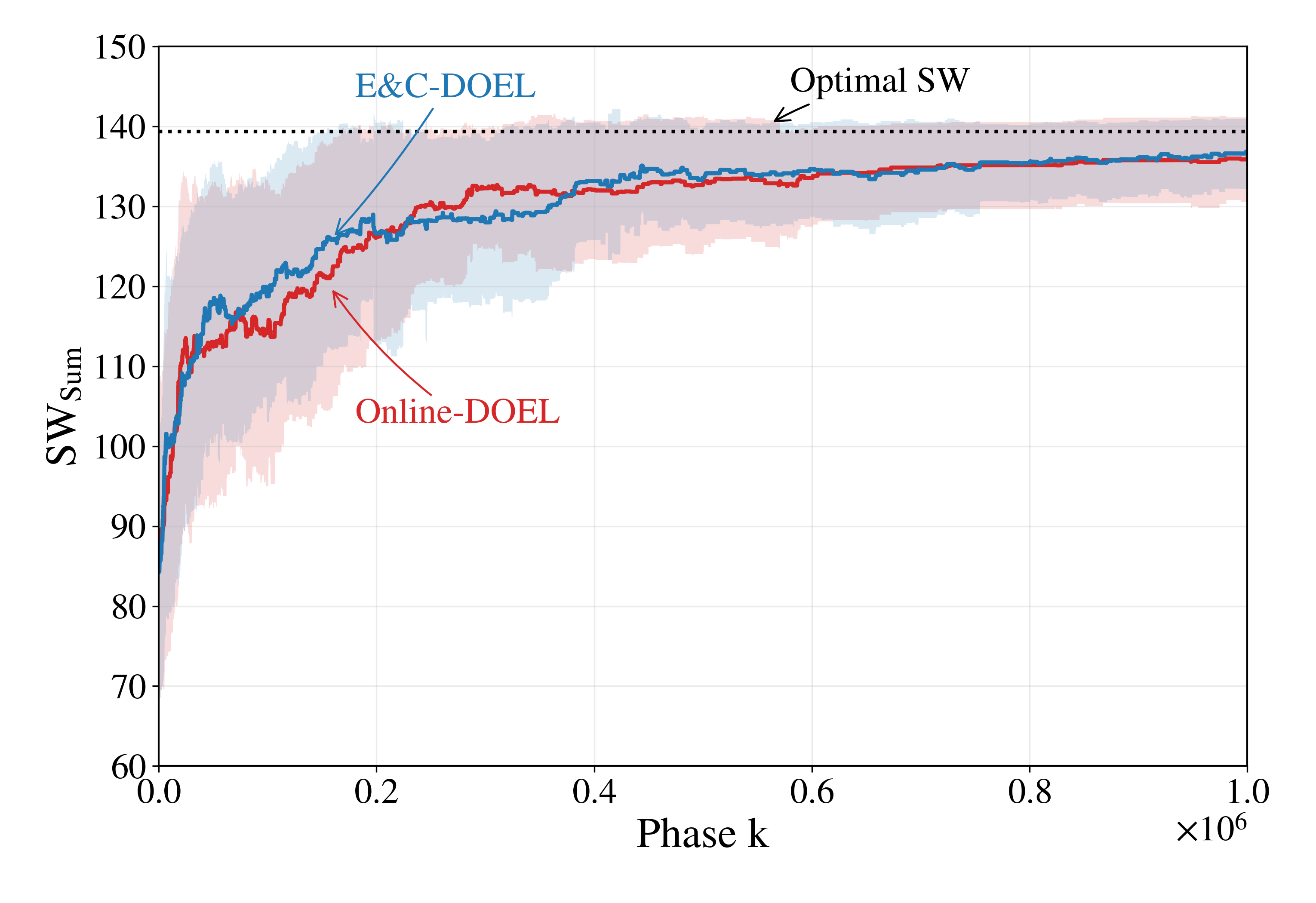}
    \caption{Sum of Local Utilities as Social Welfare}
    \label{fig:alpha0-globalmax}
  \end{subfigure}\hfill
  \begin{subfigure}{0.48\textwidth}
    \includegraphics[width=\linewidth]{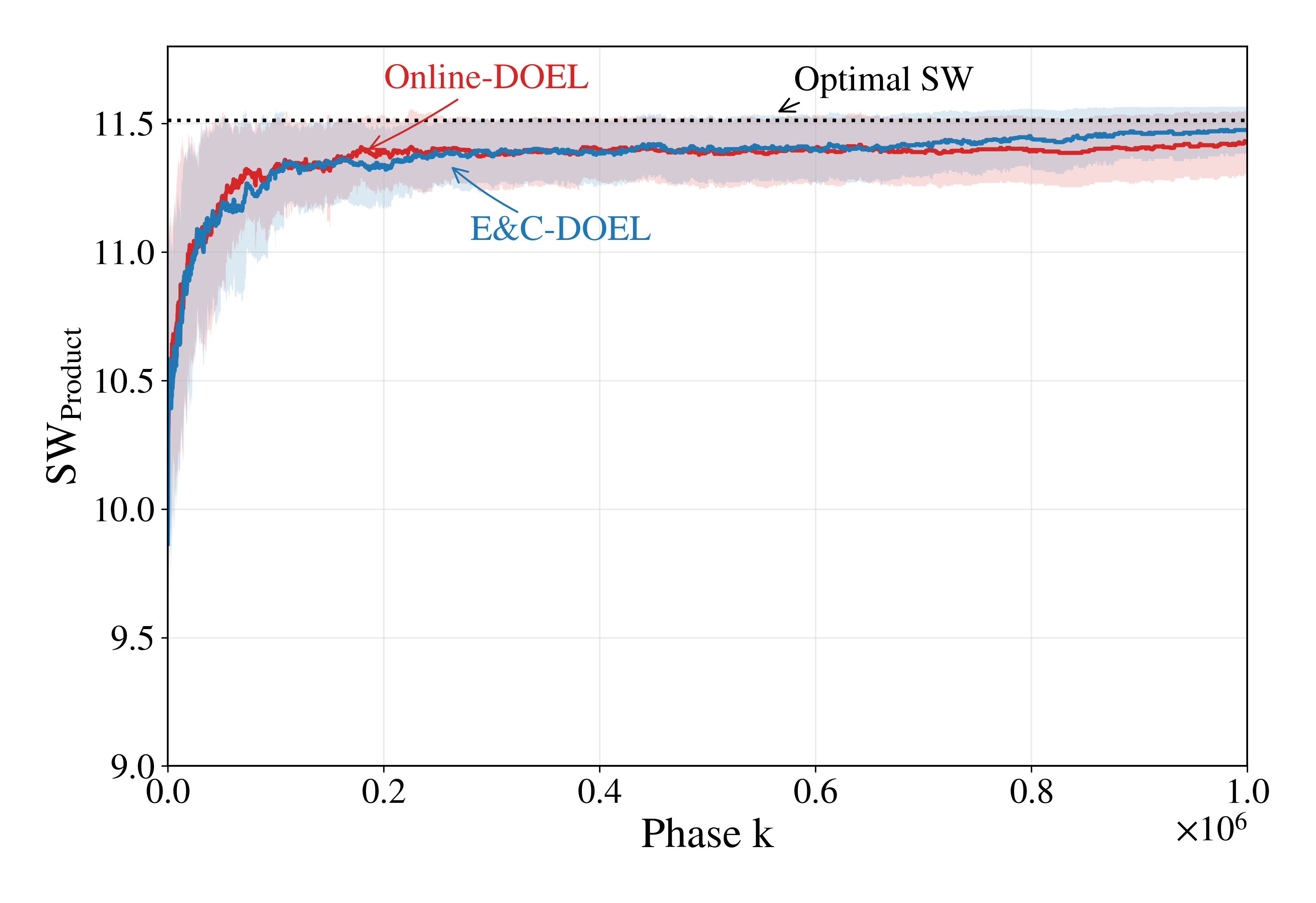}
    \caption{Product of Local Utilities as Social Welfare}
    \label{fig:alpha1-globalmax}
  \end{subfigure}
  \caption{Evolution of social welfare for content-endorsed joint strategy $\cpi_k$ under welfare-maximization.}
  \label{fig:max}
\end{figure*}

\begin{figure*}[!t]
  \centering
  \begin{subfigure}{0.48\textwidth}
    \includegraphics[width=\linewidth]{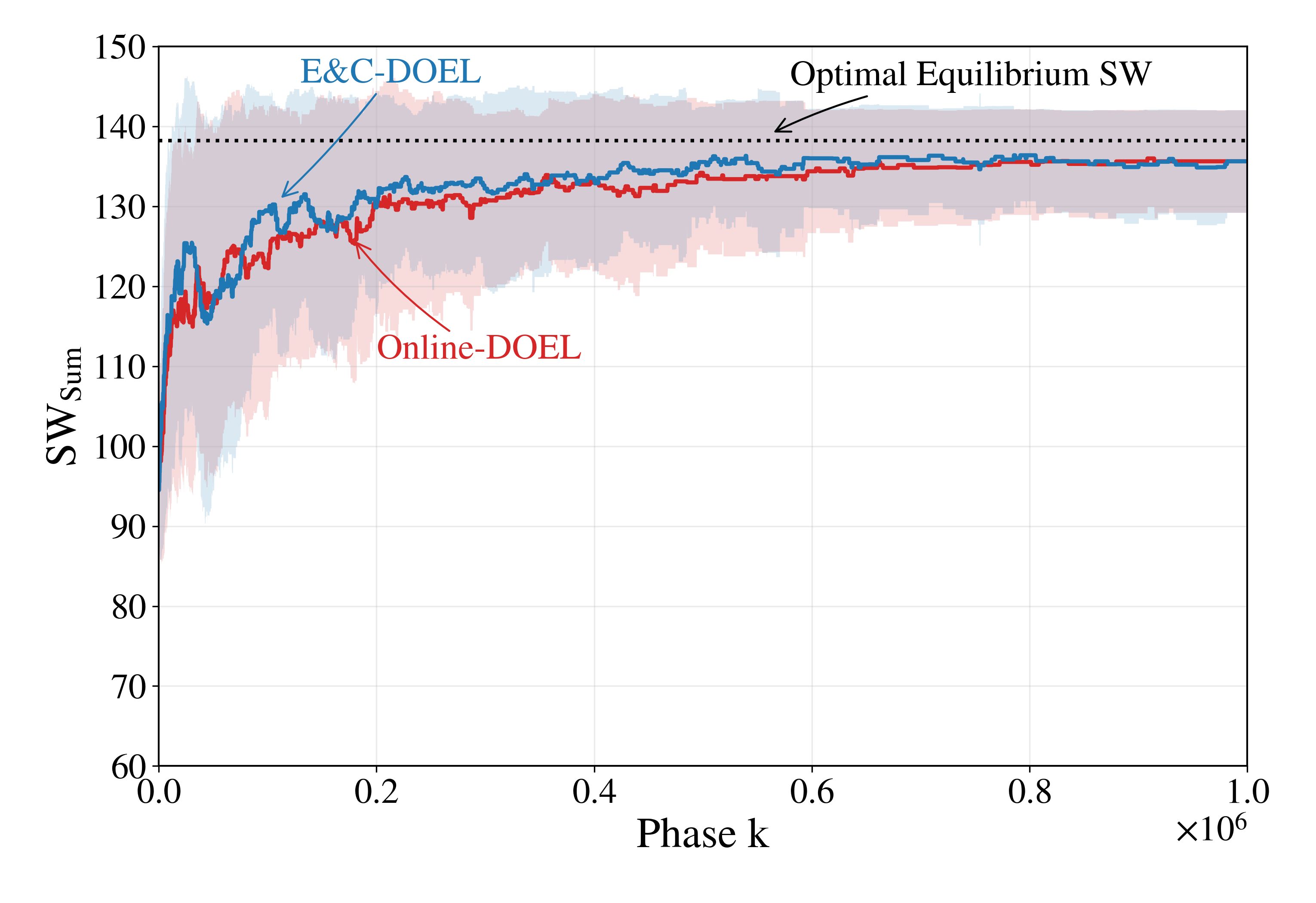}
    \caption{Sum of Local Utilities as Social Welfare}
    \label{fig:alpha0-opteq}
  \end{subfigure}\hfill
  \begin{subfigure}{0.48\textwidth}
    \includegraphics[width=\linewidth]{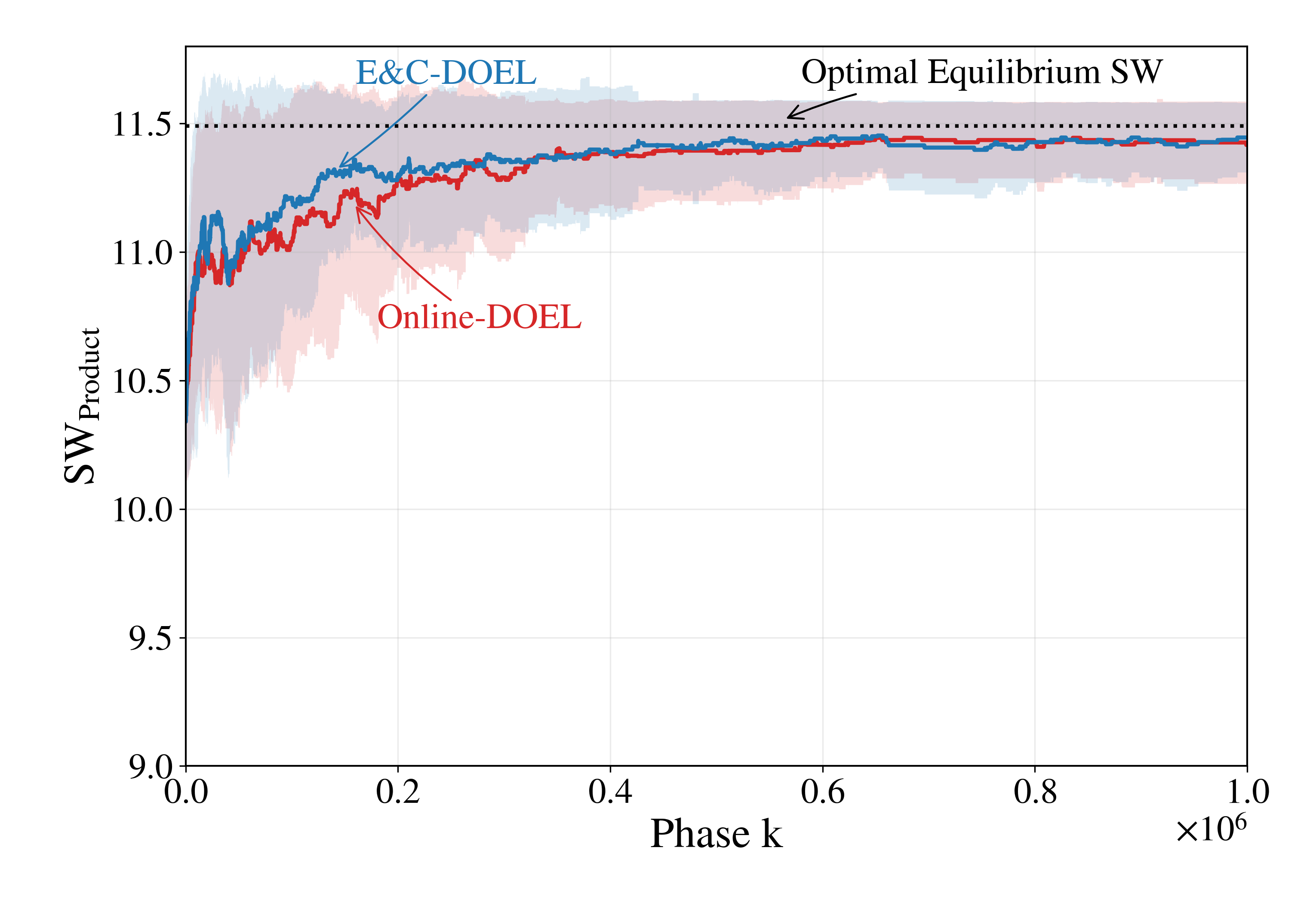}
    \caption{Product of Local Utilities as Social Welfare}
    \label{fig:alpha1-opteq}
  \end{subfigure}
  \caption{Evolution of social welfare for content-endorsed joint strategy $\cpi_k$ under equilibrium selection.}
  \label{fig:equ}
\end{figure*}

\section{Illustrative Examples}
\label{sec:simulations}

We numerically analyze the convergence behavior of Algorithm \ref{alg:main} - E\&C-DOEL and Algoritmh \ref{alg:online} - Online-DOEL for decentralized optimization and decentralized equilibrium selection with respect to the sum and product of local utilities as social welfares in $(3\times 3 \times 3)$ stochastic games, i.e., with $|N|=3$ agents, $|S|=3$ states, and $|A^i|=3$ actions per agent. Correspondingly, the joint (pure stationary) strategy space has the size of $|\Pi|=3^{3^3}=19{,}683$. The discount factor is $\gamma=0.8$. We let agents solve \eqref{eq:FPeval} and \eqref{eq:FPopt} via $120$ value iterations on the empirical MDP induced by opponents' stationary play within each exploration phase.

In Figs. \ref{fig:max} and \ref{fig:equ}, we plot the evolution of the social welfare for content-endorsed joint strategy $\cpi_k = \{\cpi^i_k \in \argmax_{\pi^i\in\Pi^i} c^i_k(\pi^i)\}_{i\in N}$ showing the welfare if each agent $i$ has committed to play $\cpi^i_k$ at phase $k$. We run $50$ independent runs. Here, the curves correspond to the average welfare across these runs with shaded areas depict $\pm 1$ standard deviation. 

For the welfare-maximization example (i.e., $\epsilon^i\rightarrow\infty$ in \eqref{eq:best}), we randomly generate rewards so that there are (i) a \emph{unique} global welfare maximizer over $\Pi$ and (ii) a visible \emph{sacrifice effect}, meaning there exists at least one agent that can attain a substantially higher individual discounted return under some other joint policy even though the resulting joint welfare is lower. This setup explicitly tests the coordination capability of the single-bit signal, ensuring agents are not merely greedily optimizing local rewards. We disable exploration perturbation, i.e., $\varepsilon^i=0$, as agents do not look for the best response, and set phase length $K=200$. We run E\&C-DOEL with $\kappa=10^6$ exploration phases. Online-DOEL runs for $10^6$ phases with $\rho_k^i=\min\{1,c/(nk)\}$, $c=350{,}000$, so there are around $6.2\times10^5$ exploration phases on the average.

In Fig.~\ref{fig:alpha0-globalmax}, $\SW_\mathrm{Sum}(\cpi_k)$, as described in \eqref{eq:SWsum}, increases rapidly early on and then approaches a stable plateau close to the dotted \textit{social optimum}. Because the plotted quantity is the welfare of the content-endorsed strategy profile $\cpi_k$, this behavior indicates that the identity of the most-frequently content-endorsed strategies becomes progressively less volatile: as $k$ grows, the counters $c^i_k(\cdot)$ concentrate on those local strategies that participate in high-welfare joint profiles, and the induced joint profile $\cpi_k$ increasingly matches (or nearly matches) the unique welfare maximizer. The separation between E\&C-DOEL and Online-DOEL is most visible in the transient: Online-DOEL's interleaving of exploration and exploitation makes $\cpi_k$ update more gradually, whereas E\&C-DOEL's pure exploration up to $\kappa$ yields faster accumulation of content events and a more decisive stabilization of $\cpi_k$. Here, we set $C^i=50$ and $\xi=(10^{-5})^{1/50}$.

In Fig.~\ref{fig:alpha1-globalmax}, $\SW_\mathrm{Product}(\cpi_k)$, as described in \eqref{eq:SWproduct}, exhibit a similar qualitative trend---fast early improvement followed by refinement---while typically showing a tighter band near the optimum, consistent with the log transform penalizing imbalances in individual utilities. Here, $C^i=\log(50)$ and $\xi=(10^{-8})^{1/\log(50)}$. Since both transitions and rewards depend on the joint action, reaching high product welfare requires not only visiting favorable states but also coordinating joint actions that improve \textit{every} agents' long-run returns.

For the equilibrium-selection example, we randomly generate reward to admit multiple stationary pure-strategy equilibria and to separate welfare maximization from equilibrium selection, i.e., the optimal equilibrium under the prescribed tolerance profile is not the unconstrained welfare maximizer. In this setting, we consider limited bounded rationality with $\epsilon^i=0.001$ and enable within-phase exploration with $\varepsilon^i=0.15$, which improves state--action coverage and stabilizes the local BE/BO computations \eqref{eq:FPeval}-\eqref{eq:FPopt} in a joint-action-dependent environment. We use phase length $K=250$ and run E\&C-DOEL with $\kappa=10^6$, while Online-DOEL runs for $10^6$ phases with the same decaying schedule $c=350{,}000$. In Figs. \ref{fig:alpha0-opteq} and \ref{fig:alpha1-opteq}, $\SW_\mathrm{Sum}(\cpi_k)$ and $\SW_\mathrm{Product}(\cpi_k)$ rise toward the corresponding dotted \textit{optimal equilibrium welfare}, indicating that as content events accumulate, the counters increasingly favor local strategies that participate in the desired equilibrium profile, so that the content-endorsed strategy profile $\cpi_k$ tracks the target where global content events are less due to $\epsilon^i-$equilibrium constraint. 

We also highlight that communication burden is very low: in a fully connected directed-broadcast network with $N=4$, each broadcast round costs $N(N-1)=12$ bits. Hence E\&C-DOEL communicates only during the $\kappa=10^6$ exploration phases and uses $12\kappa=1.2\times 10^7$ bits. Online-DOEL additionally sends an explore/exploit flag every phase.

\section{Conclusion}\label{sec:conclusion}
We proposed Decentralized Optimal Equilibrium Learning (DOEL), a framework for equilibrium selection in finite discounted stochastic games under extreme information and communication constraints. Rather than targeting equilibrium convergence alone, DOEL enables decentralized agents to optimize over the equilibrium set with respect to a designer-specified social welfare objective, while allowing heterogeneous tolerance to deviations from strict best responses. The core semantic–randomized content/discontent signaling mechanism shows that global welfare objectives can be enforced implicitly using only local observations and a single randomized bit of feedback per agent per round. Building on this mechanism, we developed explore-and-commit and online learning schemes that apply to general stochastic games, accommodate heterogeneous model-based or model-free solvers, and admit explicit finite-time regret guarantees.

Several directions remain for future research. Beyond uniform exploration, adaptive and structure-aware exploration schemes may improve efficiency and practical performance. Extending DOEL to function approximation and parametric policy classes is a promising direction for large-scale applications. Finally, the single-bit signaling paradigm provides a foundation for constrained decentralized optimization and equilibrium selection, where additional system-level constraints must be satisfied under minimal communication.

\appendix

\section{Proof of Lemma \ref{lem:SW}}\label{app:SW}

\citet[Proposition 4.2]{ref:Donmez26} yields that 
\be
|U^i(\pi) - \eU^i(\pi)| = |U^i(\pi) - U^i(\pi^i,\explore^{-i}(\pi^{-i}))| \leq  H\varepsilon,
\ee
for some $H=\frac{2}{(1-\gamma)^2}\max_{(j,s,a)}|r^j(s,a)| >0$. Then, by \eqref{eq:SW}, \eqref{eq:eSW} and Assumption \ref{assm:SW}-ii, we have
\begin{flalign}
|\eSW(\pi) - \SW(\pi)| &\leq L_g\sum_{i}w^i \cdot H\varepsilon.
\end{flalign}
Therefore, we obtain
\begin{flalign}
\Xi_\varepsilon &= \max_\pi \SW(\pi) - \max_\pi \eSW(\pi) + \eSW(\epi_*) - \SW(\epi_*)\nonumber\\
&\leq \max_\pi |\SW(\pi) - \eSW(\pi)| + \eSW(\epi_*) - \SW(\epi_*)\\
&\leq \underbrace{2L_g\sum_{i}w^i \cdot H}_{=:\overline{C}}\varepsilon,\label{eq:oC}
\end{flalign}
which completes the proof.

\section{Proof of Lemma \ref{lem:distance}}\label{app:distance}
For each parameter $\nu\sim \mathcal{V}$, let
$\lambda(\nu) := \TV(\mathcal{P},\mathcal{P}_\nu)$ be the total variation distance between $\mathcal{P}$ and $\mathcal{P}_\nu$. As the mismatch indicator, we also define $D_l = \mathbb{I}_{\{z_l\neq z_l^\ideal\}}$. Then, the maximal coupling across the ideal and actual scenarios yields that
\be
\Pr(D_l = 1 \mid \nu_l) = \Pr(z_l \neq z_l^\ideal \mid \nu_l) = \lambda(\nu_l).
\ee
Taking the expectation over $\nu_l$, we have 
$\Pr(D_l = 1) = \E[\lambda(\nu_l)] =: \lambdabar$.
Since the parameters $\nu\sim \mathcal{V}$ and couplings are independent across exploration phases $l=1,2,\ldots$, the indicators $D_1,D_2,\ldots$ are i.i.d. Bernoulli random variables with success probability $\lambdabar$. Furthermore, we have
\begin{equation}
|\wtheta_\kappa(z) - \wtheta_\kappa^\ideal(z)| = \left|\frac{1}{\kappa}\sum_{l=1}^\kappa (\mathbb{I}_{\{z_l=z\}} - \mathbb{I}_{\{z_l^\ideal=z\}})\right| \leq \frac{1}{\kappa}\sum_{l=1}^\kappa D_l=:\wlambda_\kappa\label{eq:wlambdabound}
\end{equation}
due to the triangle inequality and since $\mathbb{I}_{\{z_l=z\}} - \mathbb{I}_{\{z_l^\ideal=z\}}=0$ if $z_l=z_l^\ideal$. Then, the Hoeffding inequality yields that
\be
\Pr\big(|\wlambda_\kappa -\lambdabar| \geq c\big) \leq 2 \exp(-2\kappa c^2)\quad \forall c>0.
\ee
Therefore, we have
$\Pr(\wlambda_\kappa < c + \lambdabar) \geq 1 - 2 \exp(-2\kappa c^2)$.
Then, \eqref{eq:wlambdabound} yields that
\be
\Pr\big(|\wtheta_\kappa(z) - \wtheta_\kappa^\ideal(z)| < c + \lambdabar\big) \geq 1 - 2 \exp(-2\kappa c^2).\label{eq:diff1}
\ee

Next, we characterize an upper bound on the distance $\TV(\mathcal{P},\mathcal{P}_\nu)$. Based on the definitions \eqref{eq:joint}, \eqref{eq:jointv} of $\mathcal{P}$ and $\mathcal{P}_\nu$, we have three cases: For $\pi \in \ePieps \cap \ePiepsv$, the difference is given by
\begin{equation}
|\mathcal{P}(\pi,m) - \mathcal{P}_\nu(\pi,m)| = \frac{1}{|\Pi|}\left|\xi^{C-\eSW(\pi)} - \xi^{C-\eSWv(\pi)}\right| \leq \;\frac{1}{|\Pi|} |\log\xi| |\eSWv(\pi) - \eSW(\pi)|\label{eq:casei},
\end{equation}
for all $m \in \{0,1\}$, where the inequality follows since $\xi^{C-x}$ is $|\log\xi|$-Lipschitz function of $x$ for $\xi \in (0,1)$. For $\pi \in \ePieps \Delta \ePiepsv := (\ePieps\setminus\ePiepsv)\cup (\ePiepsv\setminus\ePieps)$, the difference is given by
\begin{equation}
|\mathcal{P}(\pi,m) - \mathcal{P}_\nu(\pi,m)| = \frac{1}{|\Pi|}\quad\forall m\in \{0,1\}.\label{eq:caseii}
\end{equation}
For $\pi\notin \ePieps \cup \ePiepsv$, the difference is zero. Combining \eqref{eq:casei} and \eqref{eq:caseii} together, we obtain
\begin{flalign}
\TV(\mathcal{P},\mathcal{P}_\nu) = \frac{1}{2}\sum_{(\pi,m)}|\mathcal{P}(\pi,m) - \mathcal{P}_\nu(\pi,m)|
&\leq |\log\xi|\max_{\pi\in\Pi} |\eSWv(\pi) - \eSW(\pi)| + \frac{|\ePieps \Delta \ePiepsv|}{|\Pi|}\nonumber\\
&\leq |\log\xi|L_g\sum_i w^i \max_{\pi\in\Pi} |v^i(s_0;\pi)-\eU^i(\pi)| \nonumber\\
&\hspace{.2in}+ \frac{|\ePieps \Delta \ePiepsv|}{|\Pi|},\label{eq:dTV}
\end{flalign}
where the last inequality follows from \eqref{eq:eSW} and Assumption \ref{assm:SW}-ii.

\begin{claim}\label{claim:sim}
Under Assumptions \ref{assm:optimal}-iii and \ref{assm:accuracy}, we have $\ePieps=\ePiepsv$, i.e., $\ePieps \Delta \ePiepsv=\varnothing$, with probability at least $1-\rho$.
\end{claim}

\begin{myproof}
Given $\eta_o$ as described in Assumption \ref{assm:optimal}-iii, assume 
\begin{subequations}\label{eq:assm}
\begin{flalign}
&|v^i(s_0;\pi) - \eU^i(\pi)|\leq \eta \\
&|v^i_*(s_0;\pi) - \max_{\tilde{\pi}^i}\eU^i(\tilde{\pi}^i,\pi^{-i})|\leq \eta
\end{flalign}
\end{subequations}
for $0<\eta<\eta_o/2$. If $\pi \in \ePieps$, then we have
\begin{equation}
v^i(s_0;\pi) \stackrel{(a)}{\geq} \eU^i(\pi) -\eta \stackrel{(b)}{\geq} \max_{\tilde{\pi}^i\in\Pi^i} \eU^i(\tilde{\pi}^i,\pi^{-i}) - \epsilon^i + \eta \stackrel{(a)}{\geq} v_*^i(s_0;\pi) - \epsilon^i,
\end{equation}
where $(a)$ follows from \eqref{eq:assm} and $(b)$ holds due to $\eta<\eta_o/2$ and Assumption \ref{assm:optimal}-iii. This yields that $\pi \in \ePiepsv$, and therefore, $\ePieps \subset \ePiepsv$. On the other hand, if $\pi\notin \ePieps$, then for some agent $i$, we have
\begin{equation}
v^i(s_0;\pi) \stackrel{(a)}{\leq} \eU^i(\pi) + \eta \stackrel{(b)}{<} \max_{\tilde{\pi}^i\in\Pi^i} \eU^i(\tilde{\pi}^i,\pi^{-i}) - \epsilon^i - \eta \stackrel{(a)}{\leq} v_*^i(s_0;\pi) - \epsilon^i,
\end{equation}
where $(a)$'s follow from \eqref{eq:assm} and $(b)$ holds due to $\eta<\eta_o/2$ and Assumption \ref{assm:optimal}-iii. This yields that $\pi\notin \ePiepsv$ and, therefore, $\ePieps \supset \ePiepsv$. 

The proof is completed since $\ePieps\subset \ePiepsv$ and  $\ePieps \supset \ePiepsv$ and the assumption \eqref{eq:assm} holds with probability at least $1-\rho$ due to Assumption \ref{assm:accuracy}.
\end{myproof}

Based on Assumption \ref{assm:accuracy} and Claim \ref{claim:sim}, we can bound \eqref{eq:dTV} by 
$\lambda(\nu)\leq L \eta$ with probability at least $1-\rho$. Then, we have
\begin{equation}
\lambdabar = \E[\lambda(\nu)\mid \lambda(\nu)\leq L\eta] \Pr(\lambda(\nu)\leq L\eta) + \E[\lambda(\nu)\mid \lambda(\nu) > L\eta] \Pr(\lambda(\nu) > L\eta) \leq L \eta + \rho \label{eq:lambdabarbound}
\end{equation}
since the distance $\lambda(\nu)\in [0,1]$. By \eqref{eq:diff1} and \eqref{eq:lambdabarbound}, we have
\begin{equation}
\Pr\big(|\wtheta_\kappa(z) - \wtheta_\kappa^\ideal(z)| < c + L \eta + \rho \big) \geq 1 - 2 \exp(-2\kappa c^2),
\end{equation}
which completes the proof.

\section{Proof of Lemma \ref{lem:separation}}\label{app:separation}

The proof follows from \citet[Lemma~1 and Proposition~1]{ref:Kiremitci26}. Particularly, we have
\begin{equation}\label{eq:bound1zeta}
\wtheta_\kappa^i(\epi^i_*,1) = \sum_{\pi^{-i}} \theta\kappa(\epi_*^i,\pi^{-i},1) > \wtheta_\kappa(\epi_*,1) > \mathcal{P}(\epi_*,1)-\zeta,
\end{equation}
where the last inequality follows from \eqref{eq:zeta}. We also have
\begin{flalign}
\mathcal{P}(\epi_*,1) -|\ePieps|\zeta &\stackrel{(a)}{=} \frac{1}{|\Pi|}\xi^{C-\eSW(\epi_*)} -|\ePieps|\zeta \stackrel{(b)}{=}\frac{1}{|\Pi|}\xi^{C-\max_{\pi\neq\epi_*}\eSW(\pi)}\big(\xi^{-\Delta_{\eSW}} - \delta\big)\\
&\stackrel{(c)}{>} \frac{|\ePieps|}{|\Pi|}\xi^{C-\max_{\pi\neq\epi_*}
\eSW(\pi)}\\
&\stackrel{(d)}{>} \sum_{\pi\in \Pi\setminus\epi_*} \mathcal{P}(\pi,1),
\end{flalign}
where $(a)$ follows from \eqref{eq:separate} and \eqref{eq:joint}, $(b)$ follows from \eqref{eq:DeltaSW} and \eqref{eq:zeta}, $(c)$ is due to Assumption \ref{assm:optimal}-ii, and $(d)$ is due to \eqref{eq:joint}. Adding $(|\ePieps|-1)\zeta$ to both hand sides, we obtain
\begin{equation}
\mathcal{P}(\epi_*,1)-\zeta > \sum_{\pi\in \Pi\setminus\epi_*} (\mathcal{P}(\pi,1) + \zeta) >  \sum_{\pi\in \Pi\setminus\epi_*} \wtheta_\kappa(\pi,1) > \wtheta_\kappa^i(\tilde{\pi}^i,1)\quad\forall \tilde{\pi}^i\neq \epi_*^i,\label{eq:bound2zeta}
\end{equation}
where the second and the third inequalities follow from \eqref{eq:zeta} and \eqref{eq:marginalization}, respectively. Then, by \eqref{eq:bound1zeta} and \eqref{eq:bound2zeta}, we have
$\wtheta_\kappa^i(\epi^i_*,1) > \wtheta_\kappa^i(\tilde{\pi}^i,1)$ for all $ \tilde{\pi}^i\neq \epi_*^i$. This completes the proof.

\section{Proof of Lemma \ref{lem:sum}}\label{app:sum}
Since $\rho_k^i = \min \{1,c/(nk)\}$, we have $\sum_{l=1}^k \overline{\rho}_l $ if $k \leq c/n$ and 
\be
\sum_{l=1}^k \overline{\rho}_l = 
\floor{\frac{c}{n}} + \sum_{l= \floor{c/n} + 1}^k(1-(1-c/(nl))^n) 
\ee
if $k > c/n$. Observe that 
\be
\sum_i\rho_l^i = \frac{c}{l} \geq \overline{\rho}_l = (1-(1-c/(nl))^n) \geq \max_i\rho_l^i = \frac{c}{nl}
\ee
for $l > c/n$. Furthermore, for integers $1\leq a\leq b$, we have
\be
\log\left(\frac{b}{a}\right) \geq \sum_{k=a+1}^b \frac{1}{k} \geq \log\left(\frac{b}{a}\right) - \frac{1}{a}.
\ee
Therefore, for $k > c/n$, we obtain
\begin{equation}
\floor{\frac{c}{n}} + c\log\left(\frac{k}{\floor{c/n}}\right) \geq \sum_{l=1}^k \overline{\rho}_l 
\geq \floor{\frac{c}{n}} + \frac{c}{n}\log\left(\frac{k}{\floor{c/n}}\right) - \frac{c/n}{\floor{c/n}},
\end{equation}
which completes the proof.

\bibliographystyle{abbrvnat}

\bibliography{main}

\end{document}